\documentclass{elsart}

\usepackage{graphicx}

\usepackage{amssymb}

\begin{document}

\begin{frontmatter}
  \title{
   Measurements of 0.2 to 20 GeV/n cosmic-ray
   proton and helium spectra from 1997 through 2002
   with the BESS spectrometer
  }
\author[Kobe]{Y. Shikaze\thanksref{jaeri}},
\author[KEK]{S. Haino\corauthref{cor1}},
\corauth[cor1]{Corresponding author.}
\ead{haino@post.kek.jp}
\author[Kobe]{K. Abe\thanksref{icrr}},
\author[ISAS]{H. Fuke},
\author[NASA]{T. Hams},
\author[UM]{K. C. Kim},
\author[KEK]{Y. Makida},
\author[KEK]{S. Matsuda},
\author[NASA]{J. W. Mitchell},
\author[NASA]{A. A. Moiseev},
\author[Tokyo]{J. Nishimura},
\author[KEK]{M. Nozaki},
\author[Tokyo]{S. Orito\thanksref{ori}},
\author[NASA]{J. F. Ormes\thanksref{denver}},
\author[Tokyo]{T. Sanuki},
\author[NASA]{M. Sasaki},
\author[UM]{E. S. Seo},
\author[NASA]{R. E. Streitmatter},
\author[KEK]{J. Suzuki},
\author[KEK]{K. Tanaka},
\author[ISAS]{T. Yamagami},
\author[KEK]{A. Yamamoto},
\author[ISAS]{T. Yoshida},
\author[KEK]{K. Yoshimura}

\address[Kobe]{Kobe University, Kobe, Hyogo 657-8501, Japan}
\address[KEK]{High Energy Accelerator Research Organization (KEK),
 Tsukuba, Ibaraki 305-0801, Japan}
\address[ISAS]{Institute of Space and Astronautical Science,
 Japan Aerospace Exploration Agency (ISAS/JAXA),
 Sagamihara, Kanagawa 229-8510, Japan}
\address[NASA]{National Aeronautics and Space Administration,
 Goddard Space Flight Center (NASA/GSFC), Greenbelt, MD 20771, USA}
\address[UM]{University of Maryland, College Park, MD 20742, USA}
\address[Tokyo]{The University of Tokyo, Bunkyo, Tokyo 113-0033 Japan}

\thanks[jaeri]{Present address: Japan Atomic Energy Research Institute,
Tokai-mura, Naka-gun, Ibaraki 319-1195, Japan}
\thanks[icrr]{Present address: ICRR, The University of Tokyo, 
              Kashiwa, Chiba 227-8582, Japan}
\thanks[ori]{deceased.}
\thanks[denver]{Present address: University of Denver, 
Denver, Colorado, 80208, USA}

\begin{abstract}
We measured low energy cosmic-ray proton and helium spectra 
in the kinetic energy range 0.215 -- 21.5~GeV/n 
at different solar activities during a period from 1997 to 2002. 
The observations were carried out with the BESS spectrometer launched 
on a balloon at Lynn Lake, Canada. A calculation for the correction 
of secondary particle backgrounds from the overlying atmosphere 
was improved by using the measured spectra at small atmospheric 
depths ranging from 5 through 37~g/cm$^2$. 
The uncertainties including statistical and systematic errors 
of the obtained spectra at the top of atmosphere are 
5--7\% for protons and 6--9\% for helium nuclei 
in the energy range 0.5 -- 5~GeV/n. 

\end{abstract}

\end{frontmatter}

\section{Introduction}
Cosmic rays, charged particles from space, enter the atmosphere
at a rate of several thousands per square meter per second. 
Among the cosmic-ray particles, protons and helium nuclei 
are dominant components. 
Thus the energy spectra and absolute fluxes of these particles
constitute fundamental data for studying cosmic-ray phenomena. 
The interstellar spectra carry important information on the origins and 
propagation of cosmic rays in the Galaxy.
At low energies, observable spectra in the heliosphere are deformed 
in a manner that depends on the solar activity. 

Cosmic-ray spectra observed at balloon altitudes suffer from
a large background of secondary particles produced inside 
the residual atmosphere. 
Therefore, in order to obtain the interstellar spectra 
and to understand the solar modulation, 
precise estimation of the atmospheric effects is important. 
The secondary components of protons are estimated 
by solving coupled transport equations 
(see for example Papini {\it et al.}~\cite{papini})
or by performing Monte Carlo simulations 
assuming elementary nuclear interaction models
between cosmic rays and atmospheric nuclei.
The estimated secondary components can be checked and modified 
by comparing them with observed spectra at various atmospheric depths.
For particles below the geomagnetic cut-off rigidity, 
atmospheric cosmic-ray spectra consist 
of purely secondary particles, and therefore, cosmic-ray measurements 
at high geomagnetic cut-off rigidity at various altitudes 
are useful to improve the accuracy of the estimation 
of the atmospheric secondary particle contribution.

We report here 0.215 to 21.5 GeV/n cosmic-ray proton and helium 
spectra measured by the BESS spectrometer for five balloon flights 
from Lynn Lake, Canada during the period 1997 through 2002. We also 
report on a realistic estimation of the atmospheric proton spectrum, 
which was tuned to be consistent with the proton spectra observed 
by the BESS spectrometer at small atmospheric depths.

The low energy cosmic-ray spectra precisely measured 
at different solar activities are important to study 
the effect of solar modulation itself and also 
the cosmic phenomena observed at different solar activities. 
We already have much precise spectral data measured at balloon altitude 
or in space~\cite{ams01,cap98}. 
This work, however, uniquely provides systematic measurements performed with 
basically the same spectrometer and at the same geographical condition. 

\section{Spectrometer}
\label{sec:spec}
\begin{figure}
  \begin{center}
    \includegraphics[width=8cm]{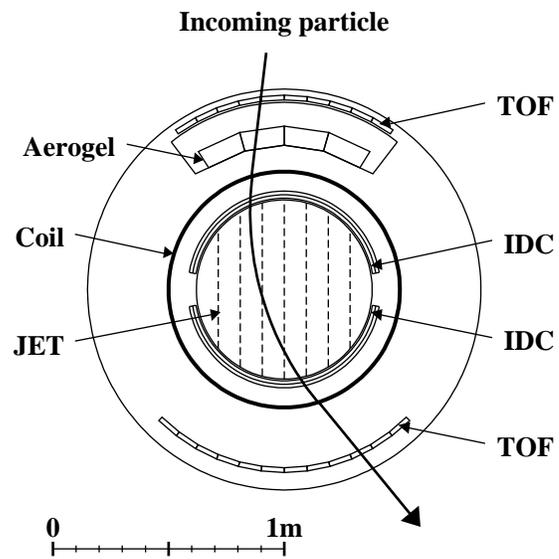}
    \caption
    {Cross-sectional view of the BESS detector in
     its 1997 configuration.}
    \label{fig:bess}
  \end{center}
\end{figure}

The detector for Balloon-borne Experiment with a Superconducting 
Spectrometer (BESS) is a high-resolution 
rigidity ($R \equiv Pc/Ze$) spectrometer with a large 
geometrical acceptance. It was designed~\cite{kn:prop87,YA88} 
and developed~\cite{YA94,Ajima,tofp,asaoka98} as an omni-purpose 
spectrometer to perform precise flux measurements 
of various components of cosmic 
rays~\cite{sanuki-phe,abe-phe,haino-phe,motoki-mu,norikura-ppb}, 
as well as highly sensitive searches for rare cosmic 
rays~\cite{YO95,MA98,OR00,MA01,AS01,book02}. 
Fig.~\ref{fig:bess} shows a schematic cross-sectional view 
of the spectrometer in its 1997 configuration.
The large acceptance is realized with a cylindrical structure 
and makes it possible to perform precise flux measurements 
with high statistics and small systematic errors. 

The spectrometer consists of a superconducting solenoid 
coil \cite{YA88}, a JET type drift chamber~(JET), 
two inner drift chambers~(IDCs), 
Time-of-Flight~(TOF) plastic scintillation hodoscopes~\cite{tofp}, 
and an aerogel \v{C}erenkov counter~\cite{asaoka98}.
A uniform magnetic field of 1~Tesla is produced by the thin 
superconducting coil which allows particles to pass
through with small interaction probability.
The magnetic field variation is less than 2.5\% along a typical trajectory.
The magnetic field region is filled with the central tracking detectors. 
Tracking of incident charged particles is performed by a circular 
fitting~\cite{Karimaki} of up to 28 hit points, 
each with a spatial resolution less than 200~$\mu$m, 
resulting in a rigidity resolution of 0.5\% at 1~GV. 
The continuous, redundant, and three-dimensional tracking makes it possible 
to recognize background events with interaction or scattering. 
The upper and lower TOF counters measure the velocity ($\beta$) 
with 1/$\beta$ resolution of 1.4\%, and provide two independent 
ionization energy loss~(d$E$/d$x$) measurements in the scintillators. 
Particle identification by mass is performed 
with these measurements($R$, $\beta$ and d$E$/d$x$). 
Furthermore a threshold-type \v{C}erenkov counter with a silica-aerogel 
radiator is installed below the upper TOF hodoscope.
The radiator, with a refractive index of 1.03 in 1997, and 1.02 
in 1998 and later, was selected to veto $e^{-}/\mu^{-}$ backgrounds for 
antiprotons up to 3.6~GeV in 1997, 
and 4.2~GeV in 1998 and later~\cite{asaoka98}. 
In the 2002 balloon-flight experiment, the observation was made with 
an upgraded detector developed for precise measurement 
of high energy particles up to 500~GeV~\cite{haino-phe,haino-nim}. 

The data acquisition sequence is initiated by a first-level TOF trigger 
which is generated by a coincidence of signals in the upper and
lower TOF hodoscopes with a threshold of more than 1/3 of 
the pulse charge for minimum ionizing particles. 
The trigger efficiency was evaluated to be 99.4 $\pm$ 0.2\% with 
a secondary proton beam at the KEK 12~GeV proton synchrotron~\cite{beamp}. 
The second-level trigger, which utilizes the hit patterns of the 
scintillator hodoscopes and the IDC for the rough rigidity determination, 
selects negatively charged particles preferentially to improve 
the statistics for the antiproton flux measurement~\cite{MA01}. 
To build an unbiased trigger sample, one of every 
60 (30 in 2000 and 10 in 2002) events is recorded. 
In 1997, 1998 and 1999, another threshold level in the TOF 
trigger was set to be at 2.5 times higher pulse 
than that from minimum ionizing particles and one of every 25 events 
was recorded to improve statistics of helium events. 
To improve statistics during balloon-ascending periods 
in 1999 and 2000, we prepared a ``low-energy proton trigger'' 
which is generated by an on-board computer 
with a sampling rate of one out of two events
to select particles with rigidity calculated to be lower 
than a fixed threshold. 
In the offline analysis, the ``low-energy proton trigger'' events 
are used to determine proton flux below 1.0~GeV during the ascent. 

In addition, an auxiliary trigger is generated 
by a signal from the \v{C}erenkov counter to record particles 
above threshold energy with sampling rates of 1/1 in 1998, 2000 and 2002, 
and 1/2 (1/1, 1/3, 1/4 during some periods) in 1999. 
This trigger improves the statistics for high energy particles. 
The efficiency of the \v{C}erenkov trigger is evaluated as the ratio 
of the \v{C}erenkov-triggered events among the unbiased trigger sample. 
It is higher than 90\% for relativistic particles ($\beta\rightarrow1$). 
For the flux determination, the \v{C}erenkov-triggered events are utilized 
above 6.31~GeV (10.0~GeV in 2002) for protons and 
3.98~GeV/n (5.41~GeV/n in 2002) for helium nuclei. 
Below these energies, the TOF-triggered events are used.

\section{Data Analysis}

\subsection{Balloon flight observations}
We carried out balloon flights to measure cosmic-ray spectra 
with the BESS spectrometer at Lynn Lake, 
Manitoba, Canada (56$^{\circ}$5'N, 101$^{\circ}$3'W), where the 
geomagnetic vertical cut-off rigidity is low, approximately 0.4~GV. 
The balloons reached a level float altitude of 37~km. The residual 
atmosphere above the spectrometer was typically around 5~g/cm$^2$. We also 
collected data during ascent in 1999, 2000 and 2002. 
A summary of the BESS balloon flights is listed in Table~\ref{tbl:flights}.

We report here primary proton and helium spectra measured at Lynn Lake 
in 1997, 1998, 1999, 2000 and 2002.
For a precise estimation of atmospheric secondary particle contributions, 
proton fluxes measured during the ascent period in 1999 and 2000, 
and during a slow descent period in 2001 were also studied. 
A detailed analysis of secondary particles 
observed at Ft. Sumner, New Mexico, USA, 
where the vertical cut off rigidity is 4.2~GV, in 2001 
is reported in Ref.~\cite{abe-phe}.

\begin{figure}
  \begin{center}
    \includegraphics[width=12cm]{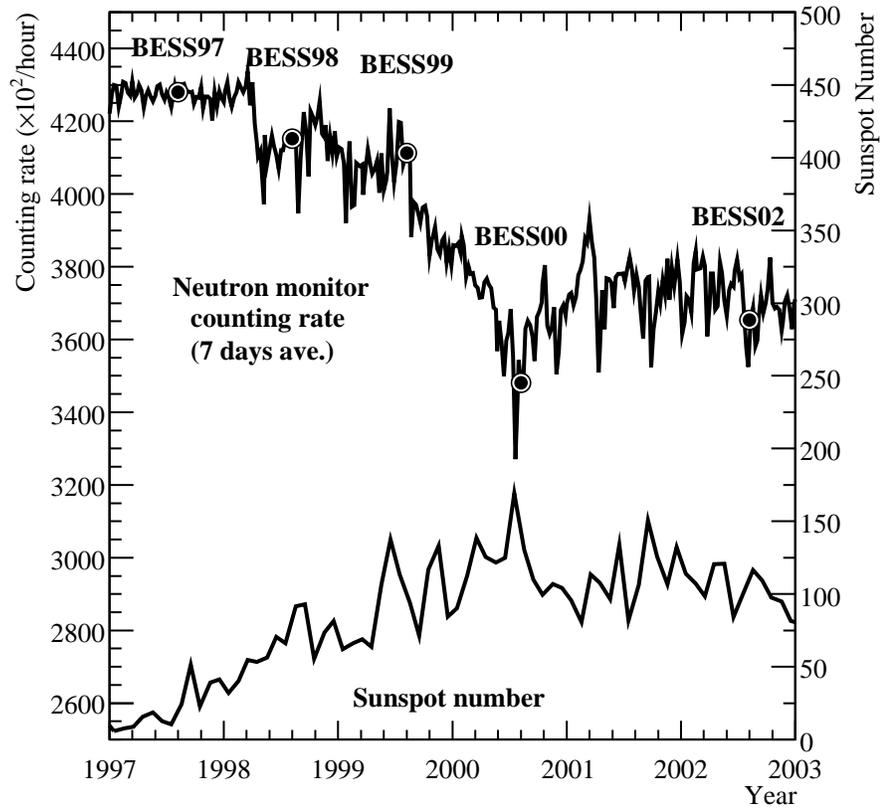}
    \caption
    {Counting rate of Climax neutron monitor (7-day average) 
     and sunspot number
     together with the BESS flight time~\cite{NEUT}.}
    \label{fig:nmbess}
  \end{center}
\end{figure}

During the period from 1997 to 2002, 
solar activity changed from the minimum to post-maximum 
as identified with sunspot number and neutron monitor 
data~\cite{NEUT} as shown in Fig.~\ref{fig:nmbess}.
The magnetic field polarity also reversed during this period.

\subsection{Data reduction}
The procedure of data analysis and derivation of the flux 
was almost the same as that of the previous balloon flight 
data~\cite{sanuki-phe,abe-phe,haino-phe,motoki-mu,norikura-ppb}, 
except for the correction of atmospheric effects.

We selected data sets which were taken with a stable condition of 
the spectrometer and flight altitude. 
The live data-taking time of the selected data sets were 
29531~sec, 25901~sec, 31890~sec, 41053~sec, and 28866~sec, 
in 1997, 1998, 1999, 2000, and 2002, respectively. 
Among the selected data sets, we first selected ``non-interacted'' 
events, i.e., those associated 
with particles passing through the detector without interaction. 
The non-interacted event was defined as an event 
which has only one isolated track, 
one or two hit-counters in each layer of the TOF hodoscopes,
and proper d$E$/d$x$ inside the upper TOF counters. 
In order to estimate efficiency of the non-interacted event selection, 
a Monte Carlo simulation with Geant3~\cite{geant} was performed. 
The probability that each particle was identified as a non-interacted event 
was evaluated by applying the same selection criteria to 
the Monte Carlo events as that applied to the observed data. 
The resultant efficiencies of non-interacted event selection 
at 0.3, 1.0 and 10~GeV/n were 82.7$\pm$0.9\%, 83.2$\pm$0.9\% 
and 76.5$\pm$1.1\%, respectively, for protons, and 78.2$\pm$1.6\%, 
72.2$\pm$1.6\% and 66.4$\pm$0.8\%, respectively, for helium nuclei. 
The systematic error was estimated by 
comparing the hit number distribution of the TOF counters.  
Based on an accelerator beam test of the BESS spectrometer~\cite{beamp}, 
the systematic uncertainty of the efficiency 
of non-interacted event selection for protons below 1~GeV should be 
within 2\%. 

The selected non-interacted events were required to pass through the  
fiducial volume defined by the central region of the JET chamber, 
TOF counters and the aerogel radiator. In the analysis of ascent data, 
the zenith angle ($\theta$) of incident particles was limited 
within $\cos \theta > 0.95$ to obtain nearly vertical flux.
In order to check the track reconstruction efficiency inside the 
tracking system, the recorded events were scanned randomly, 
and the track reconstruction efficiency was evaluated to be 99.5$\pm$0.2\%. 
It was also confirmed that the rare interacted events with more than one 
track are fully eliminated by the non-interacted event selection.

\subsection{Particle identification}
\begin{figure}
  \begin{center}
    \includegraphics[width=12cm]{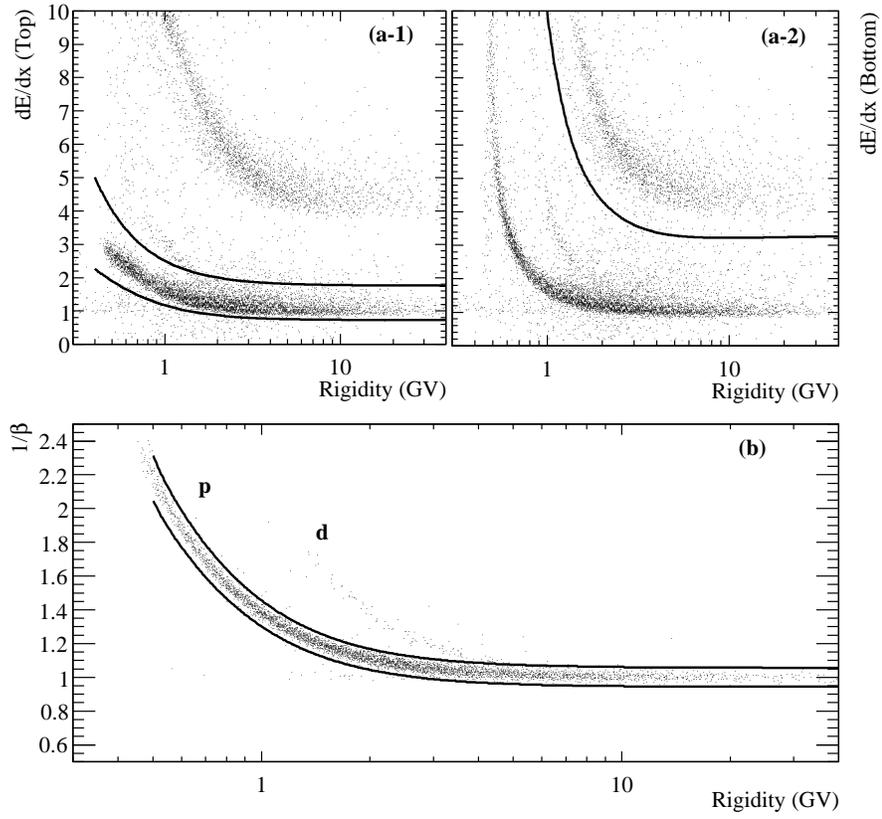}
    \caption
    {Proton selection by d$E$/d$x$-band cut for the top 
     and bottom TOF hodoscope (a-1 and a-2)
    and 1/$\beta$-band cut (b).}
    \label{fig:idplots}
  \end{center}
\end{figure}

For the non-interacted events fully contained in the fiducial region, 
particle identification was performed as follows.
The charge of the particle is identified by the ionization loss measurement.
Both d$E$/d$x$ signals from the upper and lower TOF scintillators 
were required to be proton-like. Then, particles with proton mass were 
selected by a $1/\beta$-band cut. The selection of protons with 
the d$E$/d$x$-band cut and $1/\beta$-band cut are 
shown on top and bottom in Fig.~\ref{fig:idplots}, respectively. Since at 
fixed rigidity the $1/\beta$ distribution is well described by Gaussian 
distribution and a half-width of the $1/\beta$ selection band was set 
at 3.89~$\sigma$, the selection efficiency was very close to 
unity (99.99\% for a pure Gaussian distribution).
Particle identification for helium nuclei was performed in the same way. 
However, $^3$He nuclei were included in $1/\beta$-band cut 
and were counted as helium-like events.
In conformity with previous experiments, all the helium-like events 
were treated as $^4$He nuclei in the analysis 
such as the reconstruction from rigidity to kinetic energy, 
and the efficiency estimation with the Monte Carlo simulation. 

\subsection{Quality cut}
In order to maintain the quality of the rigidity measurement, 
a track quality cut was applied to the track fitting parameters 
for high energy particles above 10~GeV/n. 
A TOF quality cut was also applied to confirm consistency between 
the hit position inside the TOF scintillator and the extrapolated track 
from the JET chamber. 

\subsection{Contamination estimation}
\subsubsection{Protons}
Protons were clearly identified 
below 1.7~GV by the mass selection as shown in Fig.~\ref{fig:idplots}.
Above 1.7~GV, however, light particles such as positrons and 
muons contaminate the $1/\beta$-band for the protons, 
and above 4~GV deuterons ($D$'s) start to contaminate it. According 
to a Monte Carlo simulation of atmospheric cosmic rays~\cite{atmnc}, 
the flux ratio of $(e^{+}+\mu^{+})/p$ is smaller than a few percent 
above 1.7~GV and decreasing with higher energy at balloon altitude. 
The observed $D$/$p$ ratio was 2\% at 3~GV by counting 
the number of events in each $1/\beta$-band cut. 
The ratio at higher energy is expected to decrease~\cite{seo} 
due to the decrease of escape path lengths of primary cosmic rays 
in the Galaxy~\cite{engelm}. 
The amount of those contaminations was as small as the statistic errors 
of the observed proton flux and no subtraction was made. 

\subsubsection{Helium nuclei}
\label{helium_contamination}
Helium nuclei were clearly identified by using both upper and lower 
TOF d$E$/d$x$ selections. 
Since the $1/\beta$-band cut includes both $^3$He and
$^4$He, the helium flux obtained includes both $^3$He and $^4$He. 
In conformity with previous experiments,
all doubly charged particles were analyzed as $^4$He.

\subsection{Corrections to the flux obtained at TOI}
The numbers of protons and helium nuclei passing through the BESS spectrometer
during the observation were obtained after correcting the detection efficiency.
Then absolute flux at the top of the instrument (TOI) 
was obtained by taking account of energy loss inside the detector, 
live time, and the geometrical acceptance. 
The energy of each particle at TOI was calculated 
by summing up the ionization energy losses inside the instrument 
by tracing back the event trajectory. The total live time of 
data-taking was measured precisely by counting 1 MHz clock 
pulses with a scaler system gated by a ``ready'' status 
that controls the first level trigger. The geometrical acceptance 
was calculated by using the simulation technique~\cite{sullivan1971}. 
In the high rigidity region 
where the bending of the track by the magnetic field is small, 
the geometrical acceptance for protons and helium nuclei is 
0.160 ${\mathrm {m^2sr}}$ without any limit for the zenith
angle for the data at the float altitude, 
while it is 0.055 ${\mathrm {m^2sr}}$ by limiting the zenith angle as 
$\cos \theta > 0.95$ for the data during the ascent period. 
The simple cylindrical shape and the uniform magnetic field facilitate
the determination of precise geometrical acceptance. 
The error arising from uncertainty of the detector alignment 
was estimated to be 1\%.

\section{Corrections of atmospheric secondary particles}
\label{sec:2ndcorr}
The flux at TOI is measured under the residual atmosphere of about 
5~g/cm$^2$, and consists of a primary component surviving without any 
nuclear interactions with air, and of a secondary component 
produced inside the overlying atmosphere. 
In order to obtain the flux at the top of the atmosphere (TOA), 
the secondary component must be estimated and subtracted. 
The secondary to primary flux ratio for protons at 0.2 GeV is as high 
as 1.0 and 0.3 during solar maximum and solar minimum, respectively, 
and the ratio decreases down to less than 3\% above 10~GeV. 
Therefore the precise estimation of the secondary component is 
essentially important to obtain low energy proton flux at TOA.
In this work we have tuned the estimation of atmospheric protons 
by using the observed data at balloon altitude as described below. 
For helium nuclei, the secondary to primary flux ratio is less than 
5\% and we used basically the same estimation as our previous 
works~\cite{sanuki-phe,haino-phe}. 

\begin{figure}
  \begin{center}
   \includegraphics[width=12cm]{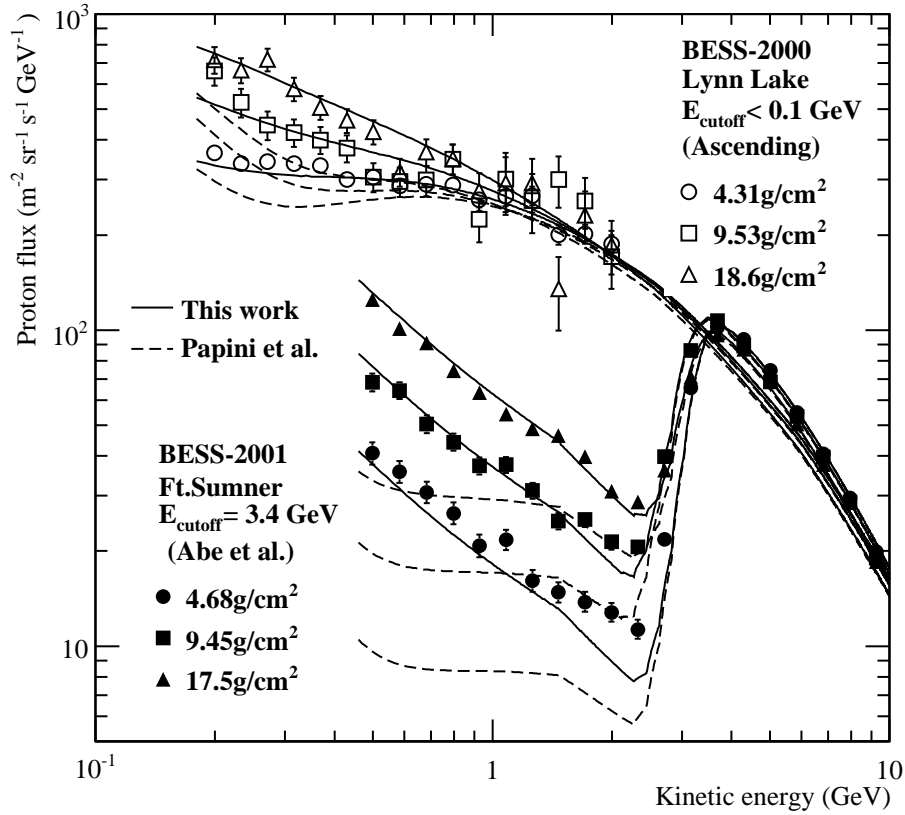}
    \caption
    {Proton energy spectra at several atmospheric depths.
     Obtained spectra measured during the ascent in 2000 
     (open markers), and the descent at Ft. Sumner 
     in 2001~\cite{abe-phe} (closed markers) are compared with
     calculated spectra before (dashed lines) and after (solid lines) 
     modification of the recoil proton production energy spectrum.}
    \label{fig:comp1}
  \end{center}
\end{figure}

\begin{figure}
  \begin{center}
    \includegraphics[width=12cm]{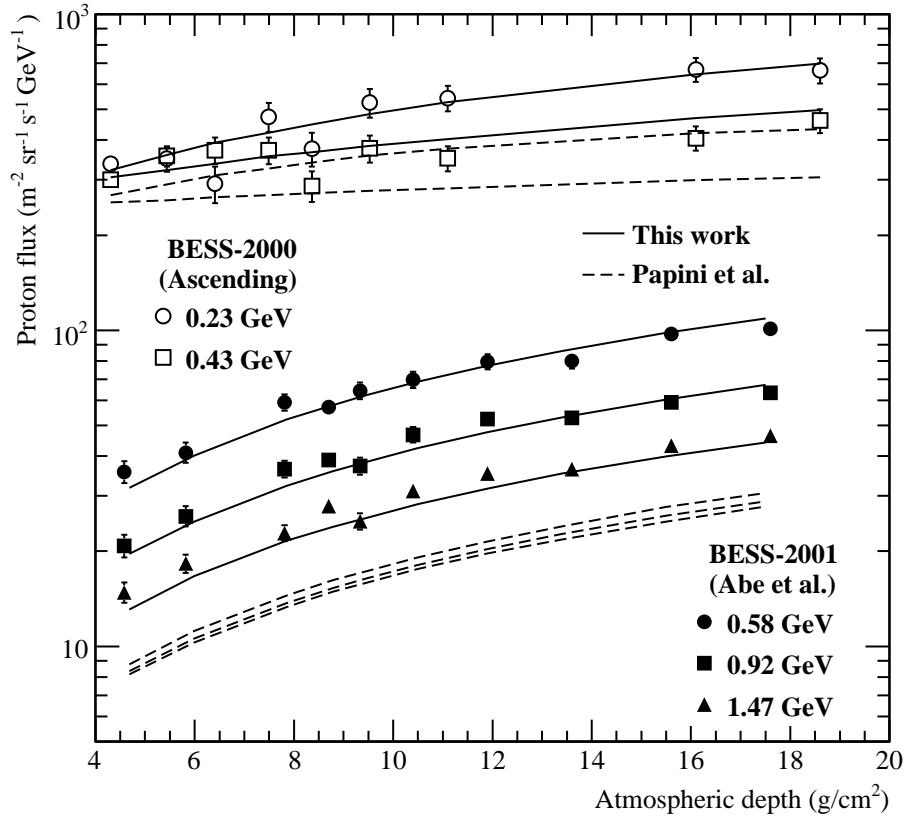}
    \caption
    {Proton flux as a function of atmospheric depth.
     Obtained fluxes measured during the ascent in 2000 
     (open markers), and the descent 
     at Ft. Sumner in 2001 (closed markers) are compared with
     calculated fluxes before (dashed lines) and after (solid lines) 
     modification of the recoil proton production energy spectrum.}
    \label{fig:comp2}
  \end{center}
\end{figure}

The proton spectrum inside the atmosphere can be estimated by solving 
simultaneous transport equations following Papini {\it et al.}~\cite{papini}. 
The primary spectrum at TOA is determined 
in an iterative procedure so that the estimated spectrum at TOI agrees 
with the observed one. 
Fig.~\ref{fig:comp1} shows the comparison between the observed 
and estimated proton spectra at several atmospheric depths 
observed during an ascent period at Lynn Lake in 2000, and
during a descent period at Ft. Sumner, New Mexico, in 2001~\cite{abe-phe}. 
Fig.~\ref{fig:comp2} shows the same comparison but as a function of 
atmospheric depth. 
At Lynn Lake, the observed spectra contain both primary and secondary 
components, but at Ft. Sumner, the spectra below the geomagnetic cut-off 
energy of 3.4~GeV should be composed of purely secondary components.
The dashed lines in Figs.~\ref{fig:comp1} and \ref{fig:comp2} 
show the estimated spectra assuming the same production spectrum 
of atmospheric secondary protons as Papini {\it et al.}~\cite{papini}.
At all the observed atmospheric depths, 
the estimated spectra are much less than the observed data. 

\begin{figure}
  \begin{center}
    \includegraphics[width=12cm]{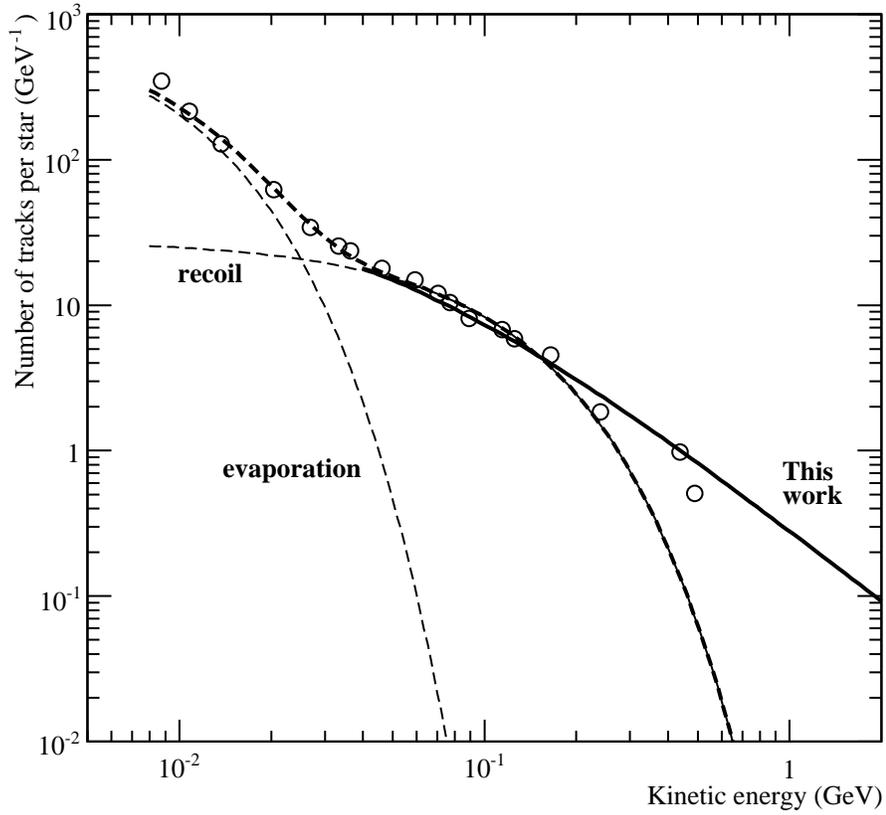}
    \caption
    {The original (dashed) and modified (solid) energy 
     spectra of secondary proton production. For the original 
     spectrum, two dominant components of evaporation and recoil 
     production from air nuclei are also shown. 
     The data points are found in Ref.~\cite{papini}}
    \label{fig:comp3}
  \end{center}
\end{figure}

In the energy region between 0.1 and a few GeV, 
the dominant source of the atmospheric secondary protons 
is recoil protons produced by interactions between 
primary cosmic rays and air nuclei. We modified the energy 
spectrum of the recoil proton production in an iterative 
procedure to reproduce the observed proton spectra at several atmospheric 
depths between 5 and 37 g/cm$^2$. The modified production spectrum 
was consistent with the original work by Papini {\it et al.} below 0.2~GeV 
as shown in Fig.~\ref{fig:comp3}. The solid lines in 
Figs.~\ref{fig:comp1} and \ref{fig:comp2} show the estimated spectra 
after the modification. The agreement between the estimation 
and the observation was significantly improved. 
We also confirmed the agreement between the estimated and observed proton 
flux during an ascent period in 1999. The relative contribution of the 
secondary component is smaller in 1999 than in 2000, 
which was during the solar maximum period. 
The modified estimation can reproduce observed data 
irrespective of the residual atmospheric depth 
or sites (Ft. Sumner and Lynn Lake).

In order to obtain the flux at the top of the atmosphere (TOA), secondary 
proton production and interaction loss of primary particles 
inside the atmosphere were estimated by solving simultaneous 
transport equations with the modified production spectrum. 
The systematic error of the atmospheric correction for proton is 
estimated as 12.7\%, which is composed of the uncertainty of 
residual air depth (9.0\%) and the cross section of primary cosmic rays 
with air nuclei (8.9\%). The systematic error on the resultant proton flux 
is estimated by multiplying the atmospheric correction factor 
as 3.7\% and 7.6\% in 1997 and 2000, respectively, at 0.2~GeV. 
The systematic error becomes smaller for higher energy and is 
1.2\% and 1.6\% in 1997 and 2000, respectively, at 1~GeV. 
The systematic error of atmospheric correction is as small as 
1\% for helium flux since the correction factor is less than 5\%.

\section{Results and Discussions}
\subsection{Proton and Helium spectra at the top of atmosphere(TOA)}
\begin{figure}
  \begin{center}
    \includegraphics[width=12cm]{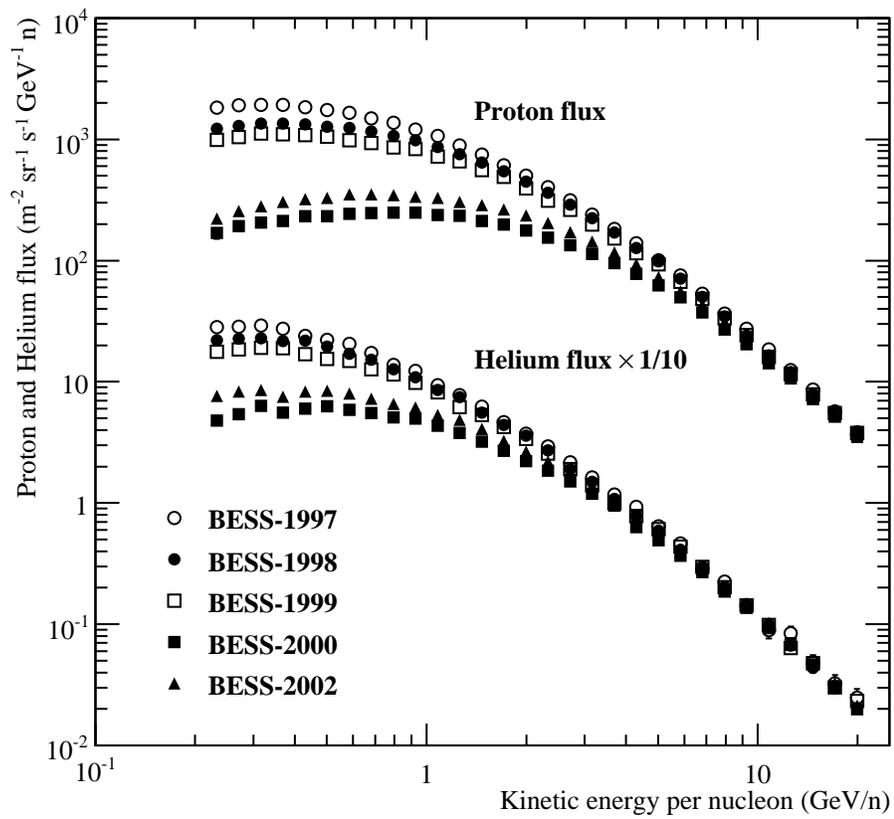}
    \caption
    {Proton and helium fluxes at the top of the atmosphere from 1997 through 2002.
    Note that helium fluxes are plotted after being divided by a factor of 10.}
    \label{fig:pheflux}
  \end{center}
\end{figure}

We obtained proton and helium spectra at the top of atmosphere 
during periods of various solar activity from 1997 to 2002 
as shown in Fig.~\ref{fig:pheflux}. The numerical data are summarized in 
Tables~\ref{tab:sump1} and \ref{tab:sumhe1}. 
The overall uncertainties of the proton spectrum including 
statistical and systematic errors are 7.8\% and 12.3\% 
in 1997 and 2000, respectively, at 0.2 GeV, 
and 7.5\% and 6.2\% in 1997 and 2000, respectively, at 20 GeV. 
The overall uncertainties of the helium spectrum are 6.3\% and 10.3\% 
in 1997 and 2000, respectively, at 0.2 GeV/n, 
and 18.2\% and 7.7\% in 1997 and 2000, respectively, at 20 GeV/n.
The difference of the uncertainties at 0.2 GeV/n is due to 
the difference of the solar activity. 
The large uncertainties of helium spectrum at 20 GeV/n in 1997 
are due to the limited statistics. As described in Section~\ref{sec:spec}, 
the \v{C}erenkov-trigger was introduced from 1998 and the statistics 
were improved. 
As written in Section~\ref{helium_contamination}, 
all doubly charged particles were analyzed as $^4$He to obtain helium flux. 
However the flux data contains a significant amount of $^3$He. 
The isotope ratio ($^3$He/$^4$He) should be 
10 -- 25~\% depending on energy~\cite{wang}.

The proton spectra at small atmospheric depths 
measured during the ascent period in 1999 and 2000 
are summarized in Tables~\ref{tab:sump3a} and \ref{tab:sump3b} 
and in Tables~\ref{tab:sump2a} and \ref{tab:sump2b}, respectively. 
These data combined with the descent data in 2001~\cite{abe-phe} 
were used to confirm our atmospheric corrections. 

A large solar modulation effect was found both 
in the proton and helium spectra in 2000 as a sudden decrease. 
On the other hand, proton and helium flux in 2002 slightly 
increased from those in 2000. We observed that the change 
in the flux variation between 2000 and 2002 was smaller than that 
between 1999 and 2000 around a period of the solar magnetic field reversal. 

Since we have used an improved method for making the atmospheric 
corrections, this analysis is improved and should replace that done 
in our previous papers BESS-1998~\cite{sanuki-phe}, 
-1999 and -2000~\cite{AS01}, and -2002~\cite{haino-phe}. 
The resultant spectra, however,  
are consistent with our previous ones above 10~GeV/n 
where the atmospheric correction is less than 3\%. 

\subsection{Force Field Approximation}
In the Force Field approximation~\cite{FFA},
cosmic-ray spectra at various solar activities are described 
by introducing the interstellar (IS) proton spectrum and 
one ``modulation parameter'' $\phi$. In this model, 
solar wind is assumed to be spherically symmetrical and is described 
with a simplified diffusion co-efficient and a transport equation
for the propagation of cosmic-ray charged particles. 
The Force Field approximation predicts 
that the cosmic-ray particle of charge $Z$ loses an energy of $Z\phi$,
as if the particle decelerated in the static field with potential of $\phi$.
Although the model is too simple to describe real dynamic activity 
of the solar wind, it has been generally used 
for the analysis of cosmic-ray spectra subject to solar modulation.
The model is convenient to describe the basic feature of modulated 
cosmic-ray spectra and to indicate a degree of the solar activity.

\begin{figure}
  \begin{center}
    \includegraphics[width=12cm]{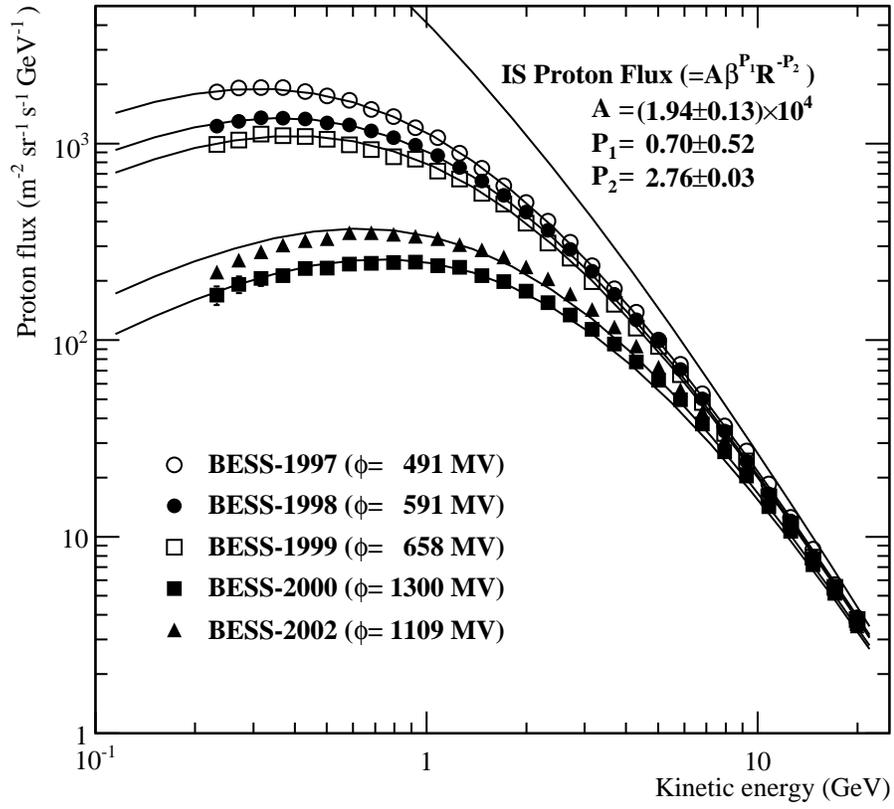}
    \caption
    {Proton spectra derived under the Force Field approximation. 
    The interstellar (IS) 
     proton spectrum was estimated by assuming $\phi$$\sim$600~MV for 
     BESS-1998. Other curves and the value of $\phi$ were obtained by 
     fitting the BESS data.}
    \label{fig:pfluxmod}
  \end{center}
\end{figure}

\begin{figure}
  \begin{center}
    \includegraphics[width=12cm]{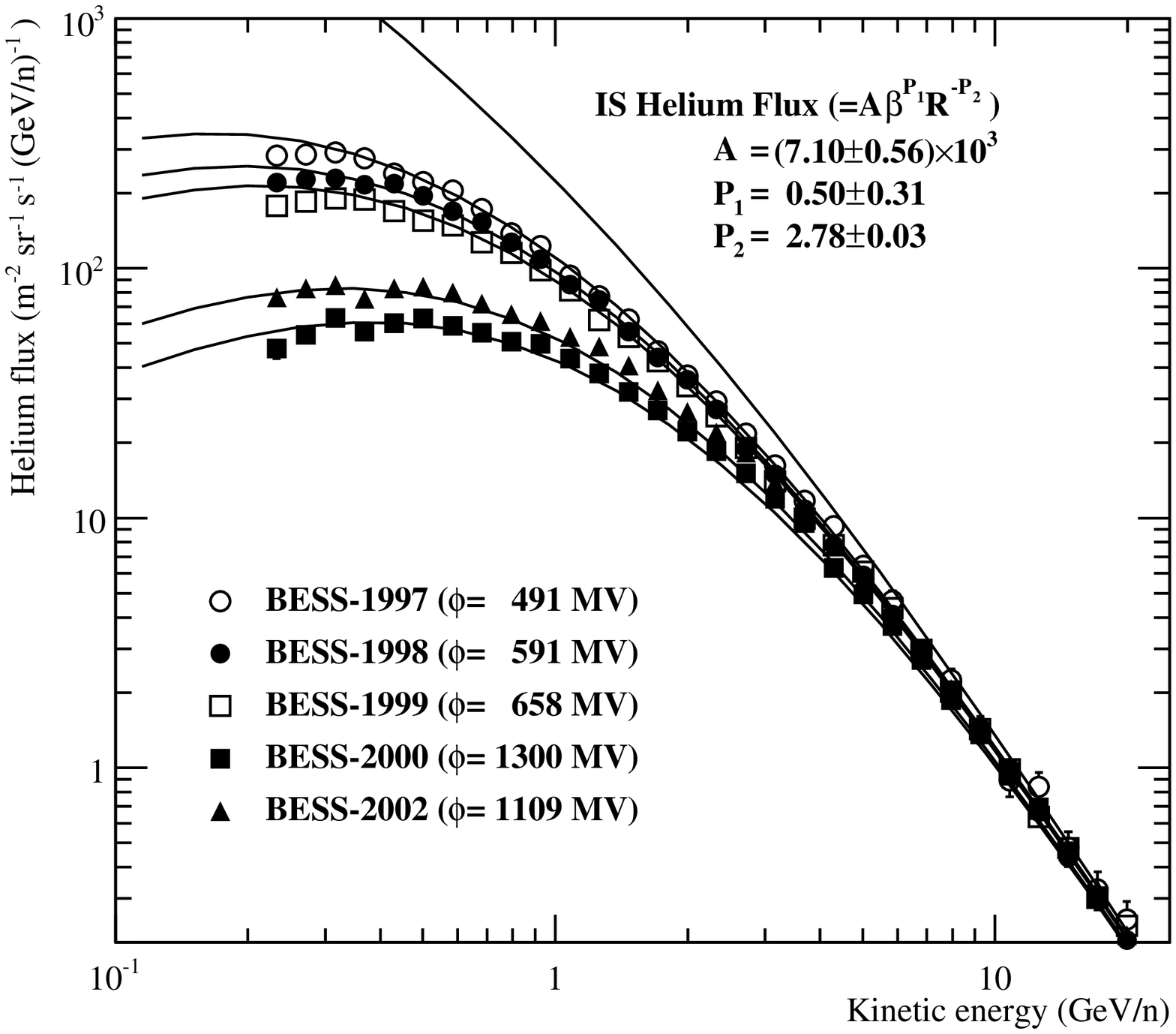}
    \caption
    {Helium spectra derived under the Force Field approximation. 
    The interstellar (IS) 
     helium spectrum was estimated by assuming $\phi$$\sim$600~MV for 
     BESS-1998. Other curves and the value of $\phi$ were obtained by 
     fitting the BESS data.}
    \label{fig:hfluxmod}
  \end{center}
\end{figure}

The proton and helium spectra obtained under this approximation are given 
in Fig.~\ref{fig:pfluxmod} and Fig.~\ref{fig:hfluxmod}, respectively,  
in comparison with our measured data. 
The IS proton spectrum was assumed to be described with 
$A \beta^{P_{1}} R^{-P_{2}}$, where, as usual, $\beta$ is the 
velocity of the particle divided by speed of light, 
$R$ is the rigidity, and $A$, $P_1$ and $P_2$ are the fitting parameters. 
The parameter $\phi$ for BESS-1998 was estimated to be $\sim$600~MV 
by Myers {\it et al.}~\cite{Myers}. 
Other curves and values of $\phi$ given in Fig.~\ref{fig:pfluxmod} were 
obtained by fitting the measured spectra using the common IS proton spectrum. 
The Force Field approximation fits relatively well to the spectra measured 
in 1997, 1998 and 1999, which are in the positive phase of the Sun's magnetic 
field polarity. However, small discrepancies can be seen in 2000 and 2002, 
which are in the negative phase. 
%
According to recent works~\cite{bieber,miyake}, the drift pattern of charged 
particles coming into the heliosphere varies with the sign 
of the polarity of heliospheric magnetic field. 
This feature cannot be treated in the Force Field approximation. 
Furthermore, the amount of energy loss depends on the observed 
particle energy, especially for negative polarity phase~\cite{miyake}. 
The small discrepancy seen in 2000 and 2002 may come 
from an inadequacy of the assumption of the model 
that the energy loss is independent of energy. 
We need further improvement in the models before we can  
consistently and accurately use them to estimate the shape of the  
interstellar spectrum.

\section{Conclusion}
Low energy cosmic-ray proton and helium spectra have been measured 
in the kinetic energy range 0.215 -- 21.5~GeV/n 
with the BESS balloon flights in northern Canada during a period of 
solar minimum, 1997, through post-maximum, 2002. 
The proton spectra at TOA in the five flight experiments were obtained 
after correction of the atmospheric proton fluxes measured in 1999 and 2000. 
The correction was further improved 
by using the pure atmospheric proton flux measured in BESS-2001 
at Ft. Sumner, where the geomagnetic cut-off energy is 3.4~GeV.
The helium spectra at TOA were obtained in the same way.
The overall uncertainties including statistical and systematic errors 
of the obtained spectra are 5--7\% for protons and 6--9\% for helium nuclei 
in the energy range 0.5 -- 5~GeV/n. The maximum uncertainties 
are 12\% for protons and 18\% for helium nuclei in the whole energy 
range of the measurements, 0.2 -- 20~GeV/n. 

Assuming an interstellar (IS) proton spectrum 
and the modulation parameter $\phi$ of about 600~MV,
the Force Field approximation can reproduce
the measured spectra obtained from the BESS-1998 experiment, 
which are in the positive phase of the Sun's magnetic field polarity. 
Further corrections or model improvements, however, 
would be required to reproduce the spectra 
during the negative polarity phase. 
The low energy cosmic-ray proton and helium spectra and the solar modulation 
effects have been much better understood based on the measurements with 
the BESS spectrometer in the solar minimum through post-maximum period. 
Those precisely measured data are important not only to study the effect 
of solar modulation but also to make a precise analysis of 
the cosmic phenomena observed at different solar activities.

\begin{ack}
We thank the NASA/GSFC/WFF Balloon office and the NSBF for the balloon 
expedition, and KEK and ISAS for various support. 
We are indebted to S.Yanagita and S.Miyake of Ibaraki University 
for their kindest cooperation of theoretical interpretations. 
This work was supported in Japan by Grant-in-Aid for Scientific
Research, MEXT and by Heiwa Nakajima Foundation and JSPS in Japan, 
and by NASA in the USA.
Analysis was performed using the computing facilities at
ICEPP, the University of Tokyo. 
\end{ack}

\clearpage

%
%
\begin{table}
\caption{A summary of the BESS flights.}
\label{tbl:flights}
\rotatebox{90}{
\begin{tabular}{lcccccccccc}
\hline
Year &1997 &1998 &1999 &2000 &2001 &2002\\
\hline
Location &Canada &Canada &Canada &Canada &U.S. &Canada\\
 &Lynn Lake &Lynn Lake &Lynn Lake &Lynn Lake &Ft.Sumner &Lynn Lake\\
Cut-off Rigidity [GV]  &0.4  &0.4  &0.4  &0.4  &4.2  &0.4\\
Launching date   &27th July &29th July &11th Aug. 
                 &10th Aug. &24th Sept. &7th Aug.\\
Float time [hr]      &20.5  &22.0  &34.5  &44.5  &14.3$^{*}$ &16.5\\
Observation time [hr]   &18.3  &20.0  &31.3  &32.5  &11.6  &13.9\\
(during ascent)         &      &      &(2.8) &(2.5) &(2.4) &(3.9)\\
Atmos. Depth [g/cm$^2$] &5.0   &5.5   &4.3   &4.3 &4.5--28 &4.6\\
Reference &\cite{MA01} &\cite{sanuki-phe} &\cite{AS01} 
          &\cite{AS01} &\cite{abe-phe} &\cite{haino-phe}\\
\hline
\multicolumn{7}{r}{$^{*}$ Including 11.8 hours of slow descent period}
\end{tabular}
}
\end{table}
\clearpage
\begin{table}
\tiny{
  \caption{Energy spectrum of protons at the top of the atmosphere.}
  \label{tab:sump1}
\rotatebox{90}{
    \begin{tabular}{clllll}
      \hline
      \begin{tabular}{@{}c@{}}       
      Energy range\\
      (GeV nucleon$^{-1}$)
      \end{tabular}
      & 
      \multicolumn{5}{c}{
      \begin{tabular}{@{}c@{}}       
	Flux$\pm~\Delta$Flux$_{\rm sta}
	\pm~\Delta$Flux$_{\rm sys}$\\
	(m$^{-2}$sr$^{-1}$s$^{-1}$GeV$^{-1}$nucleon)
      \end{tabular}
      }
      \\
      \cline{2-6}
      &       \multicolumn{5}{c}{Name of BESS flights [observed year]}\\
      &  \multicolumn{1}{c}{BESS-1997}  
      &  \multicolumn{1}{c}{BESS-1998}  
      &  \multicolumn{1}{c}{BESS-1999}  
      &  \multicolumn{1}{c}{BESS-2000}  
      &  \multicolumn{1}{c}{BESS-2002}  \\
      & \multicolumn{1}{c}{[1997]} & \multicolumn{1}{c}{[1998]} 
      & \multicolumn{1}{c}{[1999]} & \multicolumn{1}{c}{[2000]} 
      & \multicolumn{1}{c}{[2002]}  \\
      \hline 
0.215-- 0.251
& 1.83 ${\pm 0.03 \pm 0.13 \times 10^3}$
& 1.22 ${\pm 0.02 \pm 0.09 \times 10^3}$
& 9.88 ${\pm 0.20 \pm 0.76 \times 10^2}$
& 1.69 ${\pm 0.04 \pm 0.18 \times 10^2}$
& 2.22 ${\pm 0.05 \pm 0.17 \times 10^2}$
\\
0.251-- 0.293
& 1.91 ${\pm 0.02 \pm 0.13 \times 10^3}$
& 1.29 ${\pm 0.02 \pm 0.09 \times 10^3}$
& 1.04 ${\pm 0.02 \pm 0.07 \times 10^3}$
& 1.92 ${\pm 0.04 \pm 0.18 \times 10^2}$
& 2.55 ${\pm 0.05 \pm 0.17 \times 10^2}$
\\
0.293-- 0.342
& 1.92 ${\pm 0.02 \pm 0.12 \times 10^3}$
& 1.35 ${\pm 0.02 \pm 0.09 \times 10^3}$
& 1.12 ${\pm 0.02 \pm 0.07 \times 10^3}$
& 2.06 ${\pm 0.04 \pm 0.17 \times 10^2}$
& 2.81 ${\pm 0.05 \pm 0.17 \times 10^2}$
\\
0.342-- 0.398
& 1.92 ${\pm 0.02 \pm 0.12 \times 10^3}$
& 1.35 ${\pm 0.02 \pm 0.09 \times 10^3}$
& 1.10 ${\pm 0.02 \pm 0.07 \times 10^3}$
& 2.13 ${\pm 0.04 \pm 0.17 \times 10^2}$
& 3.04 ${\pm 0.05 \pm 0.16 \times 10^2}$
\\
0.398-- 0.464
& 1.84 ${\pm 0.02 \pm 0.11 \times 10^3}$
& 1.33 ${\pm 0.02 \pm 0.08 \times 10^3}$
& 1.09 ${\pm 0.02 \pm 0.07 \times 10^3}$
& 2.32 ${\pm 0.04 \pm 0.17 \times 10^2}$
& 3.21 ${\pm 0.05 \pm 0.16 \times 10^2}$
\\
0.464-- 0.541
& 1.75 ${\pm 0.02 \pm 0.10 \times 10^3}$
& 1.27 ${\pm 0.02 \pm 0.08 \times 10^3}$
& 1.05 ${\pm 0.01 \pm 0.06 \times 10^3}$
& 2.33 ${\pm 0.04 \pm 0.16 \times 10^2}$
& 3.28 ${\pm 0.04 \pm 0.15 \times 10^2}$
\\
0.541-- 0.631
& 1.66 ${\pm 0.02 \pm 0.09 \times 10^3}$
& 1.24 ${\pm 0.01 \pm 0.07 \times 10^3}$
& 9.85 ${\pm 0.13 \pm 0.57 \times 10^2}$
& 2.43 ${\pm 0.04 \pm 0.16 \times 10^2}$
& 3.51 ${\pm 0.04 \pm 0.14 \times 10^2}$
\\
0.631-- 0.736
& 1.49 ${\pm 0.01 \pm 0.08 \times 10^3}$
& 1.16 ${\pm 0.01 \pm 0.07 \times 10^3}$
& 9.33 ${\pm 0.12 \pm 0.52 \times 10^2}$
& 2.46 ${\pm 0.04 \pm 0.16 \times 10^2}$
& 3.52 ${\pm 0.04 \pm 0.13 \times 10^2}$
\\
0.736-- 0.858
& 1.37 ${\pm 0.01 \pm 0.07 \times 10^3}$
& 1.07 ${\pm 0.01 \pm 0.06 \times 10^3}$
& 8.57 ${\pm 0.10 \pm 0.47 \times 10^2}$
& 2.48 ${\pm 0.03 \pm 0.15 \times 10^2}$
& 3.46 ${\pm 0.04 \pm 0.12 \times 10^2}$
\\
0.858-- 1.000
& 1.20 ${\pm 0.01 \pm 0.06 \times 10^3}$
& 9.79 ${\pm 0.10 \pm 0.52 \times 10^2}$
& 8.32 ${\pm 0.10 \pm 0.44 \times 10^2}$
& 2.49 ${\pm 0.03 \pm 0.14 \times 10^2}$
& 3.37 ${\pm 0.03 \pm 0.11 \times 10^2}$
\\
1.00-- 1.17
& 1.07 ${\pm 0.01 \pm 0.06 \times 10^3}$
& 8.67 ${\pm 0.09 \pm 0.46 \times 10^2}$
& 7.24 ${\pm 0.08 \pm 0.38 \times 10^2}$
& 2.39 ${\pm 0.03 \pm 0.13 \times 10^2}$
& 3.28 ${\pm 0.03 \pm 0.11 \times 10^2}$
\\
1.17-- 1.36
& 8.93 ${\pm 0.08 \pm 0.46 \times 10^2}$
& 7.54 ${\pm 0.08 \pm 0.40 \times 10^2}$
& 6.60 ${\pm 0.08 \pm 0.35 \times 10^2}$
& 2.34 ${\pm 0.03 \pm 0.13 \times 10^2}$
& 3.05 ${\pm 0.03 \pm 0.10 \times 10^2}$
\\
1.36-- 1.58
& 7.49 ${\pm 0.07 \pm 0.38 \times 10^2}$
& 6.45 ${\pm 0.07 \pm 0.35 \times 10^2}$
& 5.58 ${\pm 0.06 \pm 0.30 \times 10^2}$
& 2.12 ${\pm 0.02 \pm 0.12 \times 10^2}$
& 2.87 ${\pm 0.03 \pm 0.09 \times 10^2}$
\\
1.58-- 1.85
& 6.09 ${\pm 0.06 \pm 0.31 \times 10^2}$
& 5.44 ${\pm 0.06 \pm 0.29 \times 10^2}$
& 4.91 ${\pm 0.06 \pm 0.26 \times 10^2}$
& 1.98 ${\pm 0.02 \pm 0.11 \times 10^2}$
& 2.64 ${\pm 0.02 \pm 0.08 \times 10^2}$
\\
1.85-- 2.15
& 5.01 ${\pm 0.05 \pm 0.25 \times 10^2}$
& 4.47 ${\pm 0.05 \pm 0.24 \times 10^2}$
& 3.93 ${\pm 0.05 \pm 0.21 \times 10^2}$
& 1.77 ${\pm 0.02 \pm 0.10 \times 10^2}$
& 2.36 ${\pm 0.02 \pm 0.07 \times 10^2}$
\\
2.15-- 2.51
& 4.03 ${\pm 0.04 \pm 0.20 \times 10^2}$
& 3.62 ${\pm 0.04 \pm 0.20 \times 10^2}$
& 3.11 ${\pm 0.04 \pm 0.17 \times 10^2}$
& 1.55 ${\pm 0.02 \pm 0.09 \times 10^2}$
& 2.05 ${\pm 0.02 \pm 0.06 \times 10^2}$
\\
2.51-- 2.93
& 3.14 ${\pm 0.03 \pm 0.16 \times 10^2}$
& 2.90 ${\pm 0.03 \pm 0.16 \times 10^2}$
& 2.61 ${\pm 0.03 \pm 0.15 \times 10^2}$
& 1.34 ${\pm 0.01 \pm 0.08 \times 10^2}$
& 1.72 ${\pm 0.01 \pm 0.05 \times 10^2}$
\\
2.93-- 3.42
& 2.39 ${\pm 0.03 \pm 0.12 \times 10^2}$
& 2.24 ${\pm 0.03 \pm 0.13 \times 10^2}$
& 1.98 ${\pm 0.03 \pm 0.11 \times 10^2}$
& 1.14 ${\pm 0.01 \pm 0.07 \times 10^2}$
& 1.43 ${\pm 0.01 \pm 0.04 \times 10^2}$
\\
3.42-- 3.98
& 1.83 ${\pm 0.02 \pm 0.09 \times 10^2}$
& 1.71 ${\pm 0.02 \pm 0.10 \times 10^2}$
& 1.52 ${\pm 0.02 \pm 0.09 \times 10^2}$
& 9.55 ${\pm 0.11 \pm 0.55 \times 10}$
& 1.17 ${\pm 0.01 \pm 0.03 \times 10^2}$
\\
3.98-- 4.64
& 1.39 ${\pm 0.02 \pm 0.07 \times 10^2}$
& 1.27 ${\pm 0.02 \pm 0.07 \times 10^2}$
& 1.15 ${\pm 0.02 \pm 0.07 \times 10^2}$
& 7.75 ${\pm 0.09 \pm 0.45 \times 10}$
& 9.34 ${\pm 0.09 \pm 0.27 \times 10}$
\\
4.64-- 5.41
& 1.00 ${\pm 0.01 \pm 0.05 \times 10^2}$
& 9.95 ${\pm 0.15 \pm 0.58 \times 10}$
& 9.28 ${\pm 0.15 \pm 0.54 \times 10}$
& 6.24 ${\pm 0.07 \pm 0.37 \times 10}$
& 7.30 ${\pm 0.07 \pm 0.21 \times 10}$
\\
5.41-- 6.31
& 7.54 ${\pm 0.11 \pm 0.37 \times 10}$
& 7.08 ${\pm 0.12 \pm 0.42 \times 10}$
& 6.68 ${\pm 0.12 \pm 0.39 \times 10}$
& 4.97 ${\pm 0.06 \pm 0.29 \times 10}$
& 5.59 ${\pm 0.06 \pm 0.16 \times 10}$
\\
6.31-- 7.36
& 5.33 ${\pm 0.13 \pm 0.26 \times 10}$
& 5.01 ${\pm 0.02 \pm 0.30 \times 10}$
& 4.82 ${\pm 0.03 \pm 0.29 \times 10}$
& 3.74 ${\pm 0.02 \pm 0.22 \times 10}$
& 4.30 ${\pm 0.05 \pm 0.12 \times 10}$
\\
7.36-- 8.58
& 3.65 ${\pm 0.10 \pm 0.18 \times 10}$
& 3.45 ${\pm 0.01 \pm 0.21 \times 10}$
& 3.34 ${\pm 0.02 \pm 0.20 \times 10}$
& 2.70 ${\pm 0.01 \pm 0.16 \times 10}$
& 3.11 ${\pm 0.04 \pm 0.09 \times 10}$
\\
8.58-- 10.0
& 2.73 ${\pm 0.08 \pm 0.13 \times 10}$
& 2.41 ${\pm 0.01 \pm 0.15 \times 10}$
& 2.43 ${\pm 0.02 \pm 0.15 \times 10}$
& 2.04 ${\pm 0.01 \pm 0.12 \times 10}$
& 2.22 ${\pm 0.03 \pm 0.06 \times 10}$
\\
10.0-- 11.7
& 1.86 ${\pm 0.06 \pm 0.09 \times 10}$
& 1.65 ${\pm 0.01 \pm 0.10 \times 10}$
& 1.60 ${\pm 0.01 \pm 0.10 \times 10}$
& 1.42 ${\pm 0.01 \pm 0.09 \times 10}$
& 1.55 ${\pm 0.01 \pm 0.05 \times 10}$
\\
11.7-- 13.6
& 1.25 ${\pm 0.05 \pm 0.06 \times 10}$
& 1.20 ${\pm 0.01 \pm 0.07 \times 10}$
& 1.16 ${\pm 0.01 \pm 0.07 \times 10}$
& 1.06 ${\pm 0.01 \pm 0.06 \times 10}$
& 1.09 ${\pm 0.01 \pm 0.04 \times 10}$
\\
13.6-- 15.8
& 8.63 ${\pm 0.37 \pm 0.42}$
& 7.96 ${\pm 0.05 \pm 0.49}$
& 7.88 ${\pm 0.07 \pm 0.48}$
& 7.22 ${\pm 0.04 \pm 0.44}$
& 7.49 ${\pm 0.04 \pm 0.27}$
\\
15.8-- 18.5
& 5.71 ${\pm 0.28 \pm 0.28}$
& 5.61 ${\pm 0.04 \pm 0.34}$
& 5.52 ${\pm 0.06 \pm 0.34}$
& 5.15 ${\pm 0.03 \pm 0.31}$
& 5.21 ${\pm 0.03 \pm 0.19}$
\\
18.5-- 21.5
& 3.68 ${\pm 0.21 \pm 0.18}$
& 3.91 ${\pm 0.03 \pm 0.24}$
& 3.79 ${\pm 0.05 \pm 0.23}$
& 3.60 ${\pm 0.03 \pm 0.22}$
& 3.53 ${\pm 0.03 \pm 0.14}$
\\
      \hline
    \end{tabular}
}
}
\end{table}
\clearpage
\begin{table}
\tiny{
  \caption
{Energy spectrum of helium nuclei at the top of the atmosphere.}
  \label{tab:sumhe1}
\rotatebox{90}{
    \begin{tabular}{clllll}
      \hline
      \begin{tabular}{@{}c@{}}       
      Energy range\\
      (GeV nucleon$^{-1}$)
      \end{tabular}
      & 
      \multicolumn{5}{c}{
      \begin{tabular}{@{}c@{}}       
	Flux$\pm~\Delta$Flux$_{\rm sta}
	\pm~\Delta$Flux$_{\rm sys}$\\
	(m$^{-2}$sr$^{-1}$s$^{-1}$GeV$^{-1}$nucleon)
      \end{tabular}
      }
      \\
      \cline{2-6}
      &       \multicolumn{5}{c}{Name of BESS flights [observed year]}\\
      &  \multicolumn{1}{c}{BESS-1997}  
      &  \multicolumn{1}{c}{BESS-1998}  
      &  \multicolumn{1}{c}{BESS-1999}  
      &  \multicolumn{1}{c}{BESS-2000}  
      &  \multicolumn{1}{c}{BESS-2002}  \\
      & \multicolumn{1}{c}{[1997]} & \multicolumn{1}{c}{[1998]} 
      & \multicolumn{1}{c}{[1999]} & \multicolumn{1}{c}{[2000]} 
      & \multicolumn{1}{c}{[2002]}  \\
      \hline
0.215-- 0.251
& 2.83 ${\pm 0.08 \pm 0.16 \times 10^2}$
& 2.21 ${\pm 0.07 \pm 0.12 \times 10^2}$
& 1.78 ${\pm 0.06 \pm 0.10 \times 10^2}$
& 4.78 ${\pm 0.34 \pm 0.27 \times 10}$
& 7.66 ${\pm 0.37 \pm 0.55 \times 10}$
\\
0.251-- 0.293
& 2.85 ${\pm 0.07 \pm 0.15 \times 10^2}$
& 2.27 ${\pm 0.07 \pm 0.12 \times 10^2}$
& 1.85 ${\pm 0.06 \pm 0.10 \times 10^2}$
& 5.41 ${\pm 0.34 \pm 0.30 \times 10}$
& 8.31 ${\pm 0.35 \pm 0.59 \times 10}$
\\
0.293-- 0.342
& 2.90 ${\pm 0.07 \pm 0.15 \times 10^2}$
& 2.29 ${\pm 0.06 \pm 0.12 \times 10^2}$
& 1.90 ${\pm 0.06 \pm 0.10 \times 10^2}$
& 6.32 ${\pm 0.34 \pm 0.34 \times 10}$
& 8.54 ${\pm 0.33 \pm 0.61 \times 10}$
\\
0.342-- 0.398
& 2.75 ${\pm 0.06 \pm 0.15 \times 10^2}$
& 2.16 ${\pm 0.06 \pm 0.12 \times 10^2}$
& 1.89 ${\pm 0.05 \pm 0.10 \times 10^2}$
& 5.57 ${\pm 0.30 \pm 0.30 \times 10}$
& 7.55 ${\pm 0.29 \pm 0.54 \times 10}$
\\
0.398-- 0.464
& 2.39 ${\pm 0.05 \pm 0.13 \times 10^2}$
& 2.18 ${\pm 0.05 \pm 0.12 \times 10^2}$
& 1.70 ${\pm 0.05 \pm 0.09 \times 10^2}$
& 6.02 ${\pm 0.29 \pm 0.32 \times 10}$
& 8.31 ${\pm 0.28 \pm 0.59 \times 10}$
\\
0.464-- 0.541
& 2.21 ${\pm 0.05 \pm 0.12 \times 10^2}$
& 1.95 ${\pm 0.05 \pm 0.10 \times 10^2}$
& 1.55 ${\pm 0.04 \pm 0.08 \times 10^2}$
& 6.30 ${\pm 0.28 \pm 0.34 \times 10}$
& 8.44 ${\pm 0.26 \pm 0.60 \times 10}$
\\
0.541-- 0.631
& 2.05 ${\pm 0.04 \pm 0.11 \times 10^2}$
& 1.70 ${\pm 0.04 \pm 0.09 \times 10^2}$
& 1.48 ${\pm 0.04 \pm 0.08 \times 10^2}$
& 5.87 ${\pm 0.25 \pm 0.32 \times 10}$
& 8.01 ${\pm 0.24 \pm 0.57 \times 10}$
\\
0.631-- 0.736
& 1.72 ${\pm 0.04 \pm 0.09 \times 10^2}$
& 1.53 ${\pm 0.04 \pm 0.08 \times 10^2}$
& 1.27 ${\pm 0.03 \pm 0.07 \times 10^2}$
& 5.52 ${\pm 0.23 \pm 0.30 \times 10}$
& 7.23 ${\pm 0.21 \pm 0.51 \times 10}$
\\
0.736-- 0.858
& 1.38 ${\pm 0.03 \pm 0.07 \times 10^2}$
& 1.27 ${\pm 0.03 \pm 0.07 \times 10^2}$
& 1.15 ${\pm 0.03 \pm 0.06 \times 10^2}$
& 5.08 ${\pm 0.20 \pm 0.27 \times 10}$
& 6.56 ${\pm 0.19 \pm 0.47 \times 10}$
\\
0.858-- 1.000
& 1.23 ${\pm 0.03 \pm 0.07 \times 10^2}$
& 1.09 ${\pm 0.03 \pm 0.06 \times 10^2}$
& 9.83 ${\pm 0.25 \pm 0.53 \times 10}$
& 4.97 ${\pm 0.19 \pm 0.27 \times 10}$
& 6.16 ${\pm 0.17 \pm 0.44 \times 10}$
\\
1.00-- 1.17
& 9.36 ${\pm 0.22 \pm 0.51 \times 10}$
& 8.60 ${\pm 0.22 \pm 0.47 \times 10}$
& 8.16 ${\pm 0.21 \pm 0.44 \times 10}$
& 4.34 ${\pm 0.16 \pm 0.24 \times 10}$
& 5.32 ${\pm 0.15 \pm 0.38 \times 10}$
\\
1.17-- 1.36
& 7.74 ${\pm 0.18 \pm 0.43 \times 10}$
& 7.47 ${\pm 0.19 \pm 0.41 \times 10}$
& 6.21 ${\pm 0.17 \pm 0.34 \times 10}$
& 3.80 ${\pm 0.14 \pm 0.21 \times 10}$
& 4.86 ${\pm 0.13 \pm 0.34 \times 10}$
\\
1.36-- 1.58
& 6.23 ${\pm 0.15 \pm 0.36 \times 10}$
& 5.57 ${\pm 0.15 \pm 0.31 \times 10}$
& 5.32 ${\pm 0.15 \pm 0.30 \times 10}$
& 3.21 ${\pm 0.12 \pm 0.18 \times 10}$
& 4.09 ${\pm 0.11 \pm 0.29 \times 10}$
\\
1.58-- 1.85
& 4.66 ${\pm 0.12 \pm 0.27 \times 10}$
& 4.40 ${\pm 0.13 \pm 0.25 \times 10}$
& 4.24 ${\pm 0.12 \pm 0.24 \times 10}$
& 2.70 ${\pm 0.10 \pm 0.15 \times 10}$
& 3.25 ${\pm 0.09 \pm 0.23 \times 10}$
\\
1.85-- 2.15
& 3.73 ${\pm 0.10 \pm 0.22 \times 10}$
& 3.58 ${\pm 0.11 \pm 0.21 \times 10}$
& 3.37 ${\pm 0.10 \pm 0.19 \times 10}$
& 2.22 ${\pm 0.09 \pm 0.13 \times 10}$
& 2.65 ${\pm 0.08 \pm 0.19 \times 10}$
\\
2.15-- 2.51
& 2.93 ${\pm 0.08 \pm 0.18 \times 10}$
& 2.72 ${\pm 0.08 \pm 0.16 \times 10}$
& 2.56 ${\pm 0.08 \pm 0.15 \times 10}$
& 1.85 ${\pm 0.07 \pm 0.11 \times 10}$
& 2.19 ${\pm 0.07 \pm 0.16 \times 10}$
\\
2.51-- 2.93
& 2.17 ${\pm 0.07 \pm 0.14 \times 10}$
& 1.90 ${\pm 0.07 \pm 0.11 \times 10}$
& 1.91 ${\pm 0.07 \pm 0.11 \times 10}$
& 1.51 ${\pm 0.06 \pm 0.09 \times 10}$
& 1.83 ${\pm 0.06 \pm 0.13 \times 10}$
\\
2.93-- 3.42
& 1.63 ${\pm 0.05 \pm 0.11 \times 10}$
& 1.49 ${\pm 0.05 \pm 0.09 \times 10}$
& 1.40 ${\pm 0.05 \pm 0.08 \times 10}$
& 1.19 ${\pm 0.05 \pm 0.07 \times 10}$
& 1.38 ${\pm 0.04 \pm 0.10 \times 10}$
\\
3.42-- 3.98
& 1.17 ${\pm 0.04 \pm 0.08 \times 10}$
& 1.09 ${\pm 0.04 \pm 0.07 \times 10}$
& 9.93 ${\pm 0.41 \pm 0.61}$
& 9.55 ${\pm 0.42 \pm 0.59}$
& 9.81 ${\pm 0.35 \pm 0.70}$
\\
3.98-- 4.64
& 9.28 ${\pm 0.51 \pm 0.63}$
& 7.80 ${\pm 0.11 \pm 0.49}$
& 7.79 ${\pm 0.16 \pm 0.49}$
& 6.32 ${\pm 0.09 \pm 0.39}$
& 7.69 ${\pm 0.29 \pm 0.54}$
\\
4.64-- 5.41
& 6.44 ${\pm 0.39 \pm 0.45}$
& 5.90 ${\pm 0.08 \pm 0.38}$
& 6.10 ${\pm 0.12 \pm 0.39}$
& 4.94 ${\pm 0.07 \pm 0.31}$
& 5.72 ${\pm 0.23 \pm 0.41}$
\\
5.41-- 6.31
& 4.66 ${\pm 0.31 \pm 0.34}$
& 4.10 ${\pm 0.06 \pm 0.27}$
& 4.37 ${\pm 0.09 \pm 0.28}$
& 3.69 ${\pm 0.06 \pm 0.24}$
& 3.98 ${\pm 0.06 \pm 0.32}$
\\
6.31-- 7.36
& 2.86 ${\pm 0.23 \pm 0.21}$
& 2.99 ${\pm 0.05 \pm 0.20}$
& 2.97 ${\pm 0.07 \pm 0.19}$
& 2.70 ${\pm 0.04 \pm 0.18}$
& 2.83 ${\pm 0.04 \pm 0.23}$
\\
7.36-- 8.58
& 2.24 ${\pm 0.19 \pm 0.17}$
& 2.03 ${\pm 0.04 \pm 0.14}$
& 2.02 ${\pm 0.05 \pm 0.13}$
& 1.86 ${\pm 0.03 \pm 0.12}$
& 2.07 ${\pm 0.04 \pm 0.17}$
\\
8.58-- 10.0
& 1.43 ${\pm 0.14 \pm 0.11}$
& 1.42 ${\pm 0.03 \pm 0.10}$
& 1.43 ${\pm 0.04 \pm 0.10}$
& 1.36 ${\pm 0.03 \pm 0.09}$
& 1.48 ${\pm 0.03 \pm 0.12}$
\\
10.0-- 11.7
& 8.92 ${\pm 1.04 \pm 0.70 \times 10^-1}$
& 9.99 ${\pm 0.23 \pm 0.68 \times 10^-1}$
& 9.85 ${\pm 0.34 \pm 0.67 \times 10^-1}$
& 9.32 ${\pm 0.22 \pm 0.63 \times 10^-1}$
& 1.03 ${\pm 0.02 \pm 0.08}$
\\
11.7-- 13.6
& 8.41 ${\pm 0.94 \pm 0.66 \times 10^-1}$
& 6.65 ${\pm 0.18 \pm 0.45 \times 10^-1}$
& 6.35 ${\pm 0.25 \pm 0.43 \times 10^-1}$
& 6.95 ${\pm 0.17 \pm 0.47 \times 10^-1}$
& 6.79 ${\pm 0.16 \pm 0.56 \times 10^-1}$
\\
13.6-- 15.8
& 4.78 ${\pm 0.66 \pm 0.38 \times 10^-1}$
& 4.37 ${\pm 0.13 \pm 0.30 \times 10^-1}$
& 4.77 ${\pm 0.20 \pm 0.33 \times 10^-1}$
& 4.63 ${\pm 0.13 \pm 0.32 \times 10^-1}$
& 4.73 ${\pm 0.12 \pm 0.39 \times 10^-1}$
\\
15.8-- 18.5
& 3.26 ${\pm 0.50 \pm 0.26 \times 10^-1}$
& 3.08 ${\pm 0.10 \pm 0.21 \times 10^-1}$
& 3.02 ${\pm 0.15 \pm 0.21 \times 10^-1}$
& 3.06 ${\pm 0.10 \pm 0.21 \times 10^-1}$
& 3.28 ${\pm 0.10 \pm 0.27 \times 10^-1}$
\\
18.5-- 21.5
& 2.47 ${\pm 0.41 \pm 0.20 \times 10^-1}$
& 2.04 ${\pm 0.08 \pm 0.14 \times 10^-1}$
& 2.33 ${\pm 0.12 \pm 0.16 \times 10^-1}$
& 1.99 ${\pm 0.07 \pm 0.14 \times 10^-1}$
& 2.14 ${\pm 0.07 \pm 0.18 \times 10^-1}$
\\
      \hline
    \end{tabular}
}
}
\end{table}
\clearpage
\begin{table}
\tiny{
  \caption
{Energy spectrum of protons at the small atmospheric depths measured in BESS-1999 (1).}
  \label{tab:sump3a}
\rotatebox{90}{
    \begin{tabular}{cllllll}
      \hline
      \begin{tabular}{@{}c@{}}       
      Energy range\\
      (GeV nucleon$^{-1}$)
      \end{tabular}
      & 
      \multicolumn{6}{c}{
      \begin{tabular}{@{}c@{}}       
	Flux$\pm~\Delta$Flux$_{\rm sta}
	\pm~\Delta$Flux$_{\rm sys}$\\
	(m$^{-2}$sr$^{-1}$s$^{-1}$GeV$^{-1}$nucleon)
      \end{tabular}
      }
      \\
      \cline{2-7}
      &       \multicolumn{6}{c}{atmospheric depth range [mean] (g/cm$^2$)}\\
      &  \multicolumn{1}{c}{5.25--5.79}  
      &  \multicolumn{1}{c}{5.79--6.68}  
      &  \multicolumn{1}{c}{6.68--7.66}  
      &  \multicolumn{1}{c}{7.66--8.63}  
      &  \multicolumn{1}{c}{8.63--9.70}  
      &  \multicolumn{1}{c}{9.70--11.7} \\
      & \multicolumn{1}{c}{[5.53]} & \multicolumn{1}{c}{[6.29]} 
      & \multicolumn{1}{c}{[6.97]} & \multicolumn{1}{c}{[8.10]} 
      & \multicolumn{1}{c}{[9.38]} & \multicolumn{1}{c}{[10.8]} \\
      \hline 
0.185-- 0.215
& 1.42 ${\pm 0.08 \pm 0.05 \times 10^3}$
& 1.54 ${\pm 0.09 \pm 0.06 \times 10^3}$
& 1.47 ${\pm 0.09 \pm 0.06 \times 10^3}$
& 1.42 ${\pm 0.08 \pm 0.05 \times 10^3}$
& 1.46 ${\pm 0.09 \pm 0.06 \times 10^3}$
& 1.52 ${\pm 0.08 \pm 0.06 \times 10^3}$
\\
0.215-- 0.251
& 1.38 ${\pm 0.07 \pm 0.05 \times 10^3}$
& 1.48 ${\pm 0.08 \pm 0.06 \times 10^3}$
& 1.37 ${\pm 0.08 \pm 0.05 \times 10^3}$
& 1.49 ${\pm 0.08 \pm 0.06 \times 10^3}$
& 1.62 ${\pm 0.09 \pm 0.06 \times 10^3}$
& 1.53 ${\pm 0.07 \pm 0.06 \times 10^3}$
\\
0.251-- 0.293
& 1.39 ${\pm 0.06 \pm 0.05 \times 10^3}$
& 1.33 ${\pm 0.07 \pm 0.05 \times 10^3}$
& 1.42 ${\pm 0.08 \pm 0.05 \times 10^3}$
& 1.53 ${\pm 0.07 \pm 0.06 \times 10^3}$
& 1.61 ${\pm 0.08 \pm 0.06 \times 10^3}$
& 1.58 ${\pm 0.07 \pm 0.06 \times 10^3}$
\\
0.293-- 0.341
& 1.38 ${\pm 0.06 \pm 0.05 \times 10^3}$
& 1.37 ${\pm 0.07 \pm 0.05 \times 10^3}$
& 1.37 ${\pm 0.07 \pm 0.05 \times 10^3}$
& 1.31 ${\pm 0.06 \pm 0.05 \times 10^3}$
& 1.54 ${\pm 0.07 \pm 0.06 \times 10^3}$
& 1.43 ${\pm 0.06 \pm 0.05 \times 10^3}$
\\
0.341-- 0.398
& 1.31 ${\pm 0.05 \pm 0.05 \times 10^3}$
& 1.23 ${\pm 0.06 \pm 0.05 \times 10^3}$
& 1.25 ${\pm 0.06 \pm 0.05 \times 10^3}$
& 1.27 ${\pm 0.06 \pm 0.05 \times 10^3}$
& 1.27 ${\pm 0.06 \pm 0.05 \times 10^3}$
& 1.35 ${\pm 0.06 \pm 0.05 \times 10^3}$
\\
0.398-- 0.464
& 1.22 ${\pm 0.05 \pm 0.05 \times 10^3}$
& 1.23 ${\pm 0.06 \pm 0.05 \times 10^3}$
& 1.20 ${\pm 0.06 \pm 0.05 \times 10^3}$
& 1.26 ${\pm 0.06 \pm 0.05 \times 10^3}$
& 1.21 ${\pm 0.06 \pm 0.05 \times 10^3}$
& 1.22 ${\pm 0.05 \pm 0.05 \times 10^3}$
\\
0.464-- 0.541
& 1.07 ${\pm 0.05 \pm 0.04 \times 10^3}$
& 1.11 ${\pm 0.05 \pm 0.04 \times 10^3}$
& 1.11 ${\pm 0.05 \pm 0.04 \times 10^3}$
& 9.85 ${\pm 0.47 \pm 0.38 \times 10^2}$
& 1.16 ${\pm 0.05 \pm 0.04 \times 10^3}$
& 1.07 ${\pm 0.05 \pm 0.04 \times 10^3}$
\\
0.541-- 0.631
& 1.07 ${\pm 0.05 \pm 0.04 \times 10^3}$
& 1.15 ${\pm 0.05 \pm 0.04 \times 10^3}$
& 1.06 ${\pm 0.05 \pm 0.04 \times 10^3}$
& 1.12 ${\pm 0.05 \pm 0.04 \times 10^3}$
& 1.10 ${\pm 0.05 \pm 0.04 \times 10^3}$
& 1.05 ${\pm 0.05 \pm 0.04 \times 10^3}$
\\
0.631-- 0.736
& 9.95 ${\pm 0.46 \pm 0.38 \times 10^2}$
& 1.00 ${\pm 0.05 \pm 0.04 \times 10^3}$
& 9.30 ${\pm 0.52 \pm 0.36 \times 10^2}$
& 1.04 ${\pm 0.05 \pm 0.04 \times 10^3}$
& 1.06 ${\pm 0.05 \pm 0.04 \times 10^3}$
& 1.06 ${\pm 0.05 \pm 0.04 \times 10^3}$
\\
0.736-- 0.858
& 8.99 ${\pm 0.47 \pm 0.34 \times 10^2}$
& 9.22 ${\pm 0.53 \pm 0.35 \times 10^2}$
& 8.68 ${\pm 0.54 \pm 0.33 \times 10^2}$
& 8.86 ${\pm 0.51 \pm 0.34 \times 10^2}$
& 8.81 ${\pm 0.54 \pm 0.34 \times 10^2}$
& 1.02 ${\pm 0.05 \pm 0.04 \times 10^3}$
\\
0.858-- 1.000
& 7.32 ${\pm 0.50 \pm 0.28 \times 10^2}$
& 7.52 ${\pm 0.56 \pm 0.29 \times 10^2}$
& 8.12 ${\pm 0.61 \pm 0.31 \times 10^2}$
& 8.85 ${\pm 0.59 \pm 0.34 \times 10^2}$
& 9.23 ${\pm 0.64 \pm 0.35 \times 10^2}$
& 6.89 ${\pm 0.48 \pm 0.26 \times 10^2}$
\\
1.00-- 1.17
& 8.14 ${\pm 1.21 \pm 0.31 \times 10^2}$
& 9.64 ${\pm 1.47 \pm 0.37 \times 10^2}$
& 7.43 ${\pm 1.36 \pm 0.29 \times 10^2}$
& 7.56 ${\pm 1.26 \pm 0.29 \times 10^2}$
& 8.96 ${\pm 1.45 \pm 0.34 \times 10^2}$
& 7.36 ${\pm 1.15 \pm 0.28 \times 10^2}$
\\
1.17-- 1.36
& 6.54 ${\pm 1.01 \pm 0.26 \times 10^2}$
& 5.60 ${\pm 1.04 \pm 0.22 \times 10^2}$
& 4.90 ${\pm 1.02 \pm 0.19 \times 10^2}$
& 5.42 ${\pm 0.99 \pm 0.21 \times 10^2}$
& 6.09 ${\pm 1.11 \pm 0.24 \times 10^2}$
& 6.64 ${\pm 1.01 \pm 0.26 \times 10^2}$
\\
1.36-- 1.58
& 4.64 ${\pm 0.78 \pm 0.19 \times 10^2}$
& 5.59 ${\pm 0.96 \pm 0.22 \times 10^2}$
& 5.44 ${\pm 0.99 \pm 0.22 \times 10^2}$
& 6.31 ${\pm 0.99 \pm 0.25 \times 10^2}$
& 4.15 ${\pm 0.85 \pm 0.17 \times 10^2}$
& 5.79 ${\pm 0.87 \pm 0.23 \times 10^2}$
\\
1.58-- 1.85
& 3.99 ${\pm 0.67 \pm 0.16 \times 10^2}$
& 6.35 ${\pm 0.95 \pm 0.26 \times 10^2}$
& 5.61 ${\pm 0.94 \pm 0.23 \times 10^2}$
& 3.70 ${\pm 0.70 \pm 0.15 \times 10^2}$
& 4.30 ${\pm 0.80 \pm 0.18 \times 10^2}$
& 4.52 ${\pm 0.71 \pm 0.18 \times 10^2}$
\\
1.85-- 2.15
& 3.85 ${\pm 0.62 \pm 0.16 \times 10^2}$
& 3.42 ${\pm 0.65 \pm 0.14 \times 10^2}$
& 4.46 ${\pm 0.78 \pm 0.19 \times 10^2}$
& 3.78 ${\pm 0.66 \pm 0.16 \times 10^2}$
& 4.37 ${\pm 0.75 \pm 0.18 \times 10^2}$
& 2.74 ${\pm 0.52 \pm 0.11 \times 10^2}$
\\
2.15-- 2.51
& 3.22 ${\pm 0.52 \pm 0.14 \times 10^2}$
& 3.05 ${\pm 0.57 \pm 0.13 \times 10^2}$
& 3.13 ${\pm 0.60 \pm 0.13 \times 10^2}$
& 2.75 ${\pm 0.52 \pm 0.12 \times 10^2}$
& 2.43 ${\pm 0.52 \pm 0.10 \times 10^2}$
& 2.52 ${\pm 0.46 \pm 0.11 \times 10^2}$
\\
2.51-- 2.93
& 2.75 ${\pm 0.45 \pm 0.12 \times 10^2}$
& 2.15 ${\pm 0.44 \pm 0.09 \times 10^2}$
& 2.28 ${\pm 0.47 \pm 0.10 \times 10^2}$
& 2.35 ${\pm 0.44 \pm 0.10 \times 10^2}$
& 2.45 ${\pm 0.48 \pm 0.11 \times 10^2}$
& 2.73 ${\pm 0.44 \pm 0.12 \times 10^2}$
\\
2.93-- 3.41
& 1.63 ${\pm 0.32 \pm 0.07 \times 10^2}$
& 2.26 ${\pm 0.42 \pm 0.10 \times 10^2}$
& 1.72 ${\pm 0.38 \pm 0.08 \times 10^2}$
& 2.69 ${\pm 0.44 \pm 0.12 \times 10^2}$
& 1.31 ${\pm 0.33 \pm 0.06 \times 10^2}$
& 1.81 ${\pm 0.34 \pm 0.08 \times 10^2}$
\\
3.41-- 3.98
& 1.78 ${\pm 0.31 \pm 0.08 \times 10^2}$
& 2.01 ${\pm 0.37 \pm 0.09 \times 10^2}$
& 1.55 ${\pm 0.34 \pm 0.07 \times 10^2}$
& 1.88 ${\pm 0.34 \pm 0.09 \times 10^2}$
& 1.41 ${\pm 0.31 \pm 0.06 \times 10^2}$
& 1.18 ${\pm 0.25 \pm 0.05 \times 10^2}$
\\
3.98-- 4.64
& 1.63 ${\pm 0.27 \pm 0.08 \times 10^2}$
& 1.09 ${\pm 0.25 \pm 0.05 \times 10^2}$
& 8.27 ${\pm 2.29 \pm 0.38 \times 10}$
& 1.40 ${\pm 0.28 \pm 0.06 \times 10^2}$
& 9.69 ${\pm 2.42 \pm 0.45 \times 10}$
& 1.29 ${\pm 0.24 \pm 0.06 \times 10^2}$
\\
4.64-- 5.41
& 9.13 ${\pm 1.90 \pm 0.43 \times 10}$
& 1.03 ${\pm 0.23 \pm 0.05 \times 10^2}$
& 6.52 ${\pm 1.88 \pm 0.31 \times 10}$
& 8.29 ${\pm 1.95 \pm 0.39 \times 10}$
& 1.03 ${\pm 0.23 \pm 0.05 \times 10^2}$
& 6.30 ${\pm 1.58 \pm 0.30 \times 10}$
\\
5.41-- 6.31
& 7.87 ${\pm 1.64 \pm 0.38 \times 10}$
& 4.66 ${\pm 1.41 \pm 0.22 \times 10}$
& 6.55 ${\pm 1.75 \pm 0.32 \times 10}$
& 5.56 ${\pm 1.49 \pm 0.27 \times 10}$
& 4.90 ${\pm 1.48 \pm 0.24 \times 10}$
& 6.11 ${\pm 1.44 \pm 0.29 \times 10}$
\\
6.31-- 7.36
& 4.56 ${\pm 0.24 \pm 0.22 \times 10}$
& 4.22 ${\pm 0.25 \pm 0.21 \times 10}$
& 3.78 ${\pm 0.25 \pm 0.19 \times 10}$
& 4.46 ${\pm 0.25 \pm 0.22 \times 10}$
& 4.50 ${\pm 0.27 \pm 0.22 \times 10}$
& 4.81 ${\pm 0.24 \pm 0.24 \times 10}$
\\
7.36-- 8.58
& 3.09 ${\pm 0.17 \pm 0.15 \times 10}$
& 3.11 ${\pm 0.19 \pm 0.15 \times 10}$
& 3.13 ${\pm 0.20 \pm 0.16 \times 10}$
& 3.33 ${\pm 0.19 \pm 0.17 \times 10}$
& 2.99 ${\pm 0.19 \pm 0.15 \times 10}$
& 2.96 ${\pm 0.17 \pm 0.15 \times 10}$
\\
8.58-- 10.0
& 2.03 ${\pm 0.13 \pm 0.10 \times 10}$
& 2.59 ${\pm 0.16 \pm 0.13 \times 10}$
& 2.03 ${\pm 0.15 \pm 0.10 \times 10}$
& 2.01 ${\pm 0.14 \pm 0.10 \times 10}$
& 2.06 ${\pm 0.15 \pm 0.11 \times 10}$
& 2.11 ${\pm 0.13 \pm 0.11 \times 10}$
\\
10.0-- 11.7
& 1.62 ${\pm 0.11 \pm 0.08 \times 10}$
& 1.61 ${\pm 0.12 \pm 0.08 \times 10}$
& 1.36 ${\pm 0.12 \pm 0.07 \times 10}$
& 1.41 ${\pm 0.11 \pm 0.07 \times 10}$
& 1.45 ${\pm 0.12 \pm 0.07 \times 10}$
& 1.56 ${\pm 0.11 \pm 0.08 \times 10}$
\\
11.7-- 13.6
& 1.15 ${\pm 0.08 \pm 0.06 \times 10}$
& 1.27 ${\pm 0.10 \pm 0.07 \times 10}$
& 1.11 ${\pm 0.10 \pm 0.06 \times 10}$
& 9.43 ${\pm 0.81 \pm 0.49}$
& 1.11 ${\pm 0.09 \pm 0.06 \times 10}$
& 9.96 ${\pm 0.77 \pm 0.51}$
\\
13.6-- 15.8
& 8.05 ${\pm 0.64 \pm 0.42}$
& 7.36 ${\pm 0.68 \pm 0.38}$
& 7.17 ${\pm 0.70 \pm 0.37}$
& 8.12 ${\pm 0.69 \pm 0.42}$
& 6.89 ${\pm 0.67 \pm 0.36}$
& 6.79 ${\pm 0.58 \pm 0.35}$
\\
15.8-- 18.5
& 4.71 ${\pm 0.45 \pm 0.24}$
& 4.56 ${\pm 0.49 \pm 0.23}$
& 3.98 ${\pm 0.48 \pm 0.21}$
& 4.67 ${\pm 0.48 \pm 0.24}$
& 5.07 ${\pm 0.53 \pm 0.26}$
& 4.71 ${\pm 0.45 \pm 0.24}$
\\
18.5-- 21.5
& 2.65 ${\pm 0.31 \pm 0.14}$
& 2.97 ${\pm 0.37 \pm 0.15}$
& 3.03 ${\pm 0.39 \pm 0.16}$
& 3.20 ${\pm 0.37 \pm 0.17}$
& 3.45 ${\pm 0.40 \pm 0.18}$
& 3.06 ${\pm 0.33 \pm 0.16}$
\\
      \hline
    \end{tabular}
}
}
\end{table}
\clearpage
\begin{table}
\tiny{
  \caption
{Energy spectrum of protons at the small atmospheric depths measured in BESS-1999 (2).}
  \label{tab:sump3b}
\rotatebox{90}{
    \begin{tabular}{cllllll}
      \hline
      \begin{tabular}{@{}c@{}}       
      Energy range\\
      (GeV nucleon$^{-1}$)
      \end{tabular}
      & 
      \multicolumn{6}{c}{
      \begin{tabular}{@{}c@{}}       
	Flux$\pm~\Delta$Flux$_{\rm sta}
	\pm~\Delta$Flux$_{\rm sys}$\\
	(m$^{-2}$sr$^{-1}$s$^{-1}$GeV$^{-1}$nucleon)
      \end{tabular}
      }
      \\
      \cline{2-7}
      &       \multicolumn{6}{c}{atmospheric depth range [mean] (g/cm$^2$)}\\
      &  \multicolumn{1}{c}{11.7--14.4}
      &  \multicolumn{1}{c}{14.4--17.7} 
      &  \multicolumn{1}{c}{17.7--20.7}   
      &  \multicolumn{1}{c}{20.7--24.9}   
      &  \multicolumn{1}{c}{24.9--30.5}  
      &  \multicolumn{1}{c}{30.5--37.2}   \\
      & \multicolumn{1}{c}{[12.8]}& \multicolumn{1}{c}{[16.2]} 
      & \multicolumn{1}{c}{[19.1]}& \multicolumn{1}{c}{[22.5]}
      & \multicolumn{1}{c}{[28.0]}& \multicolumn{1}{c}{[33.7]}\\
      \hline 
0.185-- 0.215
& 1.69 ${\pm 0.09 \pm 0.06 \times 10^3}$
& 1.84 ${\pm 0.09 \pm 0.07 \times 10^3}$
& 1.87 ${\pm 0.10 \pm 0.07 \times 10^3}$
& 2.11 ${\pm 0.09 \pm 0.08 \times 10^3}$
& 2.33 ${\pm 0.09 \pm 0.09 \times 10^3}$
& 2.24 ${\pm 0.10 \pm 0.08 \times 10^3}$
\\
0.215-- 0.251
& 1.64 ${\pm 0.08 \pm 0.06 \times 10^3}$
& 1.77 ${\pm 0.08 \pm 0.07 \times 10^3}$
& 1.79 ${\pm 0.09 \pm 0.07 \times 10^3}$
& 1.82 ${\pm 0.08 \pm 0.07 \times 10^3}$
& 2.01 ${\pm 0.08 \pm 0.08 \times 10^3}$
& 2.01 ${\pm 0.09 \pm 0.08 \times 10^3}$
\\
0.251-- 0.293
& 1.55 ${\pm 0.07 \pm 0.06 \times 10^3}$
& 1.70 ${\pm 0.08 \pm 0.06 \times 10^3}$
& 1.85 ${\pm 0.09 \pm 0.07 \times 10^3}$
& 1.78 ${\pm 0.07 \pm 0.07 \times 10^3}$
& 1.96 ${\pm 0.07 \pm 0.07 \times 10^3}$
& 2.07 ${\pm 0.08 \pm 0.08 \times 10^3}$
\\
0.293-- 0.341
& 1.38 ${\pm 0.06 \pm 0.05 \times 10^3}$
& 1.58 ${\pm 0.07 \pm 0.06 \times 10^3}$
& 1.60 ${\pm 0.07 \pm 0.06 \times 10^3}$
& 1.66 ${\pm 0.06 \pm 0.06 \times 10^3}$
& 1.76 ${\pm 0.06 \pm 0.07 \times 10^3}$
& 1.89 ${\pm 0.07 \pm 0.07 \times 10^3}$
\\
0.341-- 0.398
& 1.35 ${\pm 0.06 \pm 0.05 \times 10^3}$
& 1.43 ${\pm 0.06 \pm 0.05 \times 10^3}$
& 1.33 ${\pm 0.06 \pm 0.05 \times 10^3}$
& 1.51 ${\pm 0.06 \pm 0.06 \times 10^3}$
& 1.46 ${\pm 0.05 \pm 0.06 \times 10^3}$
& 1.49 ${\pm 0.06 \pm 0.06 \times 10^3}$
\\
0.398-- 0.464
& 1.21 ${\pm 0.05 \pm 0.05 \times 10^3}$
& 1.32 ${\pm 0.06 \pm 0.05 \times 10^3}$
& 1.27 ${\pm 0.06 \pm 0.05 \times 10^3}$
& 1.30 ${\pm 0.05 \pm 0.05 \times 10^3}$
& 1.26 ${\pm 0.05 \pm 0.05 \times 10^3}$
& 1.31 ${\pm 0.06 \pm 0.05 \times 10^3}$
\\
0.464-- 0.541
& 1.19 ${\pm 0.05 \pm 0.05 \times 10^3}$
& 1.21 ${\pm 0.05 \pm 0.05 \times 10^3}$
& 1.11 ${\pm 0.05 \pm 0.04 \times 10^3}$
& 1.16 ${\pm 0.05 \pm 0.04 \times 10^3}$
& 1.21 ${\pm 0.05 \pm 0.05 \times 10^3}$
& 1.20 ${\pm 0.05 \pm 0.05 \times 10^3}$
\\
0.541-- 0.631
& 1.08 ${\pm 0.05 \pm 0.04 \times 10^3}$
& 1.01 ${\pm 0.05 \pm 0.04 \times 10^3}$
& 1.09 ${\pm 0.05 \pm 0.04 \times 10^3}$
& 1.08 ${\pm 0.05 \pm 0.04 \times 10^3}$
& 1.12 ${\pm 0.04 \pm 0.04 \times 10^3}$
& 1.13 ${\pm 0.05 \pm 0.04 \times 10^3}$
\\
0.631-- 0.736
& 9.73 ${\pm 0.47 \pm 0.37 \times 10^2}$
& 1.06 ${\pm 0.05 \pm 0.04 \times 10^3}$
& 1.01 ${\pm 0.05 \pm 0.04 \times 10^3}$
& 9.74 ${\pm 0.45 \pm 0.37 \times 10^2}$
& 9.93 ${\pm 0.43 \pm 0.38 \times 10^2}$
& 1.07 ${\pm 0.05 \pm 0.04 \times 10^3}$
\\
0.736-- 0.858
& 9.50 ${\pm 0.51 \pm 0.36 \times 10^2}$
& 9.07 ${\pm 0.51 \pm 0.35 \times 10^2}$
& 9.48 ${\pm 0.56 \pm 0.36 \times 10^2}$
& 9.13 ${\pm 0.47 \pm 0.35 \times 10^2}$
& 1.03 ${\pm 0.05 \pm 0.04 \times 10^3}$
& 8.56 ${\pm 0.50 \pm 0.33 \times 10^2}$
\\
0.858-- 1.000
& 8.49 ${\pm 0.56 \pm 0.32 \times 10^2}$
& 8.26 ${\pm 0.57 \pm 0.31 \times 10^2}$
& 8.30 ${\pm 0.61 \pm 0.32 \times 10^2}$
& 8.85 ${\pm 0.54 \pm 0.34 \times 10^2}$
& 8.64 ${\pm 0.51 \pm 0.33 \times 10^2}$
& 7.92 ${\pm 0.56 \pm 0.30 \times 10^2}$
\\
1.00-- 1.17
& 7.00 ${\pm 1.17 \pm 0.27 \times 10^2}$
& 7.07 ${\pm 1.21 \pm 0.27 \times 10^2}$
& 5.57 ${\pm 1.16 \pm 0.21 \times 10^2}$
& 5.62 ${\pm 0.99 \pm 0.22 \times 10^2}$
& 7.18 ${\pm 1.07 \pm 0.28 \times 10^2}$
& 5.83 ${\pm 1.10 \pm 0.22 \times 10^2}$
\\
1.17-- 1.36
& 5.53 ${\pm 0.96 \pm 0.22 \times 10^2}$
& 6.98 ${\pm 1.12 \pm 0.27 \times 10^2}$
& 8.75 ${\pm 1.35 \pm 0.34 \times 10^2}$
& 5.29 ${\pm 0.89 \pm 0.21 \times 10^2}$
& 6.46 ${\pm 0.94 \pm 0.25 \times 10^2}$
& 5.73 ${\pm 1.01 \pm 0.22 \times 10^2}$
\\
1.36-- 1.58
& 4.28 ${\pm 0.78 \pm 0.17 \times 10^2}$
& 3.81 ${\pm 0.76 \pm 0.15 \times 10^2}$
& 4.26 ${\pm 0.87 \pm 0.17 \times 10^2}$
& 6.30 ${\pm 0.90 \pm 0.25 \times 10^2}$
& 4.91 ${\pm 0.76 \pm 0.20 \times 10^2}$
& 5.49 ${\pm 0.92 \pm 0.22 \times 10^2}$
\\
1.58-- 1.85
& 5.39 ${\pm 0.81 \pm 0.22 \times 10^2}$
& 4.97 ${\pm 0.81 \pm 0.20 \times 10^2}$
& 4.42 ${\pm 0.82 \pm 0.18 \times 10^2}$
& 3.65 ${\pm 0.63 \pm 0.15 \times 10^2}$
& 5.32 ${\pm 0.73 \pm 0.22 \times 10^2}$
& 3.80 ${\pm 0.71 \pm 0.16 \times 10^2}$
\\
1.85-- 2.15
& 4.24 ${\pm 0.67 \pm 0.18 \times 10^2}$
& 3.85 ${\pm 0.66 \pm 0.16 \times 10^2}$
& 2.77 ${\pm 0.60 \pm 0.12 \times 10^2}$
& 3.54 ${\pm 0.58 \pm 0.15 \times 10^2}$
& 3.31 ${\pm 0.54 \pm 0.14 \times 10^2}$
& 3.40 ${\pm 0.62 \pm 0.14 \times 10^2}$
\\
2.15-- 2.51
& 3.01 ${\pm 0.52 \pm 0.13 \times 10^2}$
& 3.12 ${\pm 0.55 \pm 0.13 \times 10^2}$
& 2.38 ${\pm 0.52 \pm 0.10 \times 10^2}$
& 2.47 ${\pm 0.45 \pm 0.10 \times 10^2}$
& 3.07 ${\pm 0.48 \pm 0.13 \times 10^2}$
& 3.32 ${\pm 0.57 \pm 0.14 \times 10^2}$
\\
2.51-- 2.93
& 3.03 ${\pm 0.49 \pm 0.13 \times 10^2}$
& 2.66 ${\pm 0.47 \pm 0.12 \times 10^2}$
& 3.29 ${\pm 0.56 \pm 0.14 \times 10^2}$
& 1.82 ${\pm 0.36 \pm 0.08 \times 10^2}$
& 1.79 ${\pm 0.34 \pm 0.08 \times 10^2}$
& 2.50 ${\pm 0.46 \pm 0.11 \times 10^2}$
\\
2.93-- 3.41
& 2.50 ${\pm 0.41 \pm 0.11 \times 10^2}$
& 1.66 ${\pm 0.35 \pm 0.07 \times 10^2}$
& 1.68 ${\pm 0.38 \pm 0.07 \times 10^2}$
& 2.31 ${\pm 0.38 \pm 0.10 \times 10^2}$
& 1.94 ${\pm 0.33 \pm 0.09 \times 10^2}$
& 1.66 ${\pm 0.35 \pm 0.07 \times 10^2}$
\\
3.41-- 3.98
& 1.45 ${\pm 0.29 \pm 0.07 \times 10^2}$
& 1.24 ${\pm 0.28 \pm 0.06 \times 10^2}$
& 1.95 ${\pm 0.38 \pm 0.09 \times 10^2}$
& 1.10 ${\pm 0.24 \pm 0.05 \times 10^2}$
& 1.43 ${\pm 0.26 \pm 0.06 \times 10^2}$
& 1.49 ${\pm 0.30 \pm 0.07 \times 10^2}$
\\
3.98-- 4.64
& 1.20 ${\pm 0.24 \pm 0.06 \times 10^2}$
& 1.01 ${\pm 0.23 \pm 0.05 \times 10^2}$
& 9.95 ${\pm 2.49 \pm 0.46 \times 10}$
& 1.49 ${\pm 0.26 \pm 0.07 \times 10^2}$
& 1.11 ${\pm 0.21 \pm 0.05 \times 10^2}$
& 1.28 ${\pm 0.26 \pm 0.06 \times 10^2}$
\\
4.64-- 5.41
& 8.11 ${\pm 1.86 \pm 0.38 \times 10}$
& 5.93 ${\pm 1.65 \pm 0.28 \times 10}$
& 1.06 ${\pm 0.24 \pm 0.05 \times 10^2}$
& 9.24 ${\pm 1.89 \pm 0.44 \times 10}$
& 7.35 ${\pm 1.60 \pm 0.35 \times 10}$
& 9.60 ${\pm 2.09 \pm 0.45 \times 10}$
\\
5.41-- 6.31
& 5.52 ${\pm 1.42 \pm 0.27 \times 10}$
& 6.68 ${\pm 1.62 \pm 0.32 \times 10}$
& 7.32 ${\pm 1.83 \pm 0.35 \times 10}$
& 7.96 ${\pm 1.63 \pm 0.38 \times 10}$
& 3.92 ${\pm 1.09 \pm 0.19 \times 10}$
& 4.72 ${\pm 1.36 \pm 0.23 \times 10}$
\\
6.31-- 7.36
& 3.90 ${\pm 0.23 \pm 0.19 \times 10}$
& 4.12 ${\pm 0.24 \pm 0.20 \times 10}$
& 3.81 ${\pm 0.25 \pm 0.19 \times 10}$
& 4.16 ${\pm 0.22 \pm 0.20 \times 10}$
& 3.80 ${\pm 0.20 \pm 0.19 \times 10}$
& 3.84 ${\pm 0.23 \pm 0.19 \times 10}$
\\
7.36-- 8.58
& 2.73 ${\pm 0.17 \pm 0.14 \times 10}$
& 2.83 ${\pm 0.18 \pm 0.14 \times 10}$
& 2.73 ${\pm 0.19 \pm 0.14 \times 10}$
& 2.79 ${\pm 0.16 \pm 0.14 \times 10}$
& 2.94 ${\pm 0.16 \pm 0.15 \times 10}$
& 2.63 ${\pm 0.17 \pm 0.13 \times 10}$
\\
8.58-- 10.0
& 2.28 ${\pm 0.14 \pm 0.12 \times 10}$
& 1.95 ${\pm 0.13 \pm 0.10 \times 10}$
& 2.04 ${\pm 0.15 \pm 0.10 \times 10}$
& 1.90 ${\pm 0.12 \pm 0.10 \times 10}$
& 1.78 ${\pm 0.11 \pm 0.09 \times 10}$
& 1.69 ${\pm 0.12 \pm 0.09 \times 10}$
\\
10.0-- 11.7
& 1.39 ${\pm 0.10 \pm 0.07 \times 10}$
& 1.60 ${\pm 0.11 \pm 0.08 \times 10}$
& 1.38 ${\pm 0.11 \pm 0.07 \times 10}$
& 1.42 ${\pm 0.10 \pm 0.07 \times 10}$
& 1.42 ${\pm 0.09 \pm 0.07 \times 10}$
& 1.17 ${\pm 0.10 \pm 0.06 \times 10}$
\\
11.7-- 13.6
& 1.11 ${\pm 0.08 \pm 0.06 \times 10}$
& 1.11 ${\pm 0.09 \pm 0.06 \times 10}$
& 1.02 ${\pm 0.09 \pm 0.05 \times 10}$
& 9.39 ${\pm 0.74 \pm 0.48}$
& 1.03 ${\pm 0.07 \pm 0.05 \times 10}$
& 8.53 ${\pm 0.77 \pm 0.44}$
\\
13.6-- 15.8
& 6.01 ${\pm 0.57 \pm 0.31}$
& 6.77 ${\pm 0.63 \pm 0.35}$
& 6.40 ${\pm 0.66 \pm 0.33}$
& 6.49 ${\pm 0.56 \pm 0.34}$
& 5.68 ${\pm 0.50 \pm 0.29}$
& 5.85 ${\pm 0.58 \pm 0.30}$
\\
15.8-- 18.5
& 4.00 ${\pm 0.43 \pm 0.21}$
& 4.33 ${\pm 0.46 \pm 0.22}$
& 3.72 ${\pm 0.46 \pm 0.19}$
& 4.48 ${\pm 0.43 \pm 0.23}$
& 4.53 ${\pm 0.41 \pm 0.23}$
& 4.68 ${\pm 0.48 \pm 0.24}$
\\
18.5-- 21.5
& 3.04 ${\pm 0.34 \pm 0.16}$
& 3.04 ${\pm 0.36 \pm 0.16}$
& 3.49 ${\pm 0.41 \pm 0.18}$
& 3.03 ${\pm 0.33 \pm 0.16}$
& 2.69 ${\pm 0.29 \pm 0.14}$
& 3.13 ${\pm 0.36 \pm 0.16}$
\\
      \hline
    \end{tabular}
}
}
\end{table}
\clearpage
\begin{table}
\tiny{
  \caption
{Energy spectrum of protons at the small atmospheric depths measured in BESS-2000 (1).}
  \label{tab:sump2a}
\rotatebox{90}{
    \begin{tabular}{cllllll}
      \hline
      \begin{tabular}{@{}c@{}}       
      Energy range\\
      (GeV nucleon$^{-1}$)
      \end{tabular}
      & 
      \multicolumn{6}{c}{
      \begin{tabular}{@{}c@{}}       
	Flux$\pm~\Delta$Flux$_{\rm sta}
	\pm~\Delta$Flux$_{\rm sys}$\\
	(m$^{-2}$sr$^{-1}$s$^{-1}$GeV$^{-1}$nucleon)
      \end{tabular}
      }
      \\
      \cline{2-7}
      &       \multicolumn{6}{c}{atmospheric depth range [mean] (g/cm$^2$)}\\
      &  \multicolumn{1}{c}{4.00--5.00}  
      &  \multicolumn{1}{c}{5.00--6.00}  
      &  \multicolumn{1}{c}{6.00--7.00}  
      &  \multicolumn{1}{c}{7.00--8.00}  
      &  \multicolumn{1}{c}{8.00--9.00}  
      &  \multicolumn{1}{c}{9.00--10.0} \\
      & \multicolumn{1}{c}{[4.31]} & \multicolumn{1}{c}{[5.44]} 
      & \multicolumn{1}{c}{[6.41]} & \multicolumn{1}{c}{[7.49]} 
      & \multicolumn{1}{c}{[8.36]} & \multicolumn{1}{c}{[9.53]} \\
      \hline 
0.185-- 0.215
& 3.63 ${\pm 0.12 \pm 0.14 \times 10^2}$
& 3.40 ${\pm 0.31 \pm 0.13 \times 10^2}$
& 4.48 ${\pm 0.50 \pm 0.17 \times 10^2}$
& 4.67 ${\pm 0.52 \pm 0.18 \times 10^2}$
& 4.31 ${\pm 0.51 \pm 0.16 \times 10^2}$
& 6.61 ${\pm 0.63 \pm 0.25 \times 10^2}$
\\
0.215-- 0.251
& 3.37 ${\pm 0.11 \pm 0.13 \times 10^2}$
& 3.49 ${\pm 0.28 \pm 0.13 \times 10^2}$
& 2.91 ${\pm 0.37 \pm 0.11 \times 10^2}$
& 4.74 ${\pm 0.47 \pm 0.18 \times 10^2}$
& 3.75 ${\pm 0.44 \pm 0.14 \times 10^2}$
& 5.25 ${\pm 0.51 \pm 0.20 \times 10^2}$
\\
0.251-- 0.293
& 3.41 ${\pm 0.10 \pm 0.13 \times 10^2}$
& 3.44 ${\pm 0.26 \pm 0.13 \times 10^2}$
& 3.56 ${\pm 0.37 \pm 0.14 \times 10^2}$
& 4.34 ${\pm 0.42 \pm 0.17 \times 10^2}$
& 4.47 ${\pm 0.44 \pm 0.17 \times 10^2}$
& 4.45 ${\pm 0.43 \pm 0.17 \times 10^2}$
\\
0.293-- 0.341
& 3.36 ${\pm 0.09 \pm 0.13 \times 10^2}$
& 3.71 ${\pm 0.25 \pm 0.14 \times 10^2}$
& 3.46 ${\pm 0.34 \pm 0.13 \times 10^2}$
& 3.48 ${\pm 0.35 \pm 0.13 \times 10^2}$
& 4.47 ${\pm 0.41 \pm 0.17 \times 10^2}$
& 4.22 ${\pm 0.39 \pm 0.16 \times 10^2}$
\\
0.341-- 0.398
& 3.31 ${\pm 0.09 \pm 0.13 \times 10^2}$
& 3.37 ${\pm 0.23 \pm 0.13 \times 10^2}$
& 4.01 ${\pm 0.35 \pm 0.15 \times 10^2}$
& 3.07 ${\pm 0.31 \pm 0.12 \times 10^2}$
& 4.09 ${\pm 0.37 \pm 0.16 \times 10^2}$
& 3.99 ${\pm 0.36 \pm 0.15 \times 10^2}$
\\
0.398-- 0.464
& 3.00 ${\pm 0.08 \pm 0.11 \times 10^2}$
& 3.56 ${\pm 0.22 \pm 0.14 \times 10^2}$
& 3.71 ${\pm 0.32 \pm 0.14 \times 10^2}$
& 3.72 ${\pm 0.33 \pm 0.14 \times 10^2}$
& 2.86 ${\pm 0.30 \pm 0.11 \times 10^2}$
& 3.76 ${\pm 0.34 \pm 0.14 \times 10^2}$
\\
0.464-- 0.541
& 3.05 ${\pm 0.08 \pm 0.12 \times 10^2}$
& 3.02 ${\pm 0.20 \pm 0.12 \times 10^2}$
& 3.44 ${\pm 0.31 \pm 0.13 \times 10^2}$
& 3.30 ${\pm 0.30 \pm 0.13 \times 10^2}$
& 3.57 ${\pm 0.33 \pm 0.14 \times 10^2}$
& 3.05 ${\pm 0.30 \pm 0.12 \times 10^2}$
\\
0.541-- 0.631
& 2.86 ${\pm 0.08 \pm 0.11 \times 10^2}$
& 3.06 ${\pm 0.20 \pm 0.12 \times 10^2}$
& 3.38 ${\pm 0.31 \pm 0.13 \times 10^2}$
& 3.44 ${\pm 0.31 \pm 0.13 \times 10^2}$
& 3.02 ${\pm 0.30 \pm 0.12 \times 10^2}$
& 2.95 ${\pm 0.29 \pm 0.11 \times 10^2}$
\\
0.631-- 0.736
& 2.90 ${\pm 0.08 \pm 0.11 \times 10^2}$
& 3.33 ${\pm 0.22 \pm 0.13 \times 10^2}$
& 2.63 ${\pm 0.28 \pm 0.10 \times 10^2}$
& 3.09 ${\pm 0.30 \pm 0.12 \times 10^2}$
& 3.15 ${\pm 0.32 \pm 0.12 \times 10^2}$
& 2.98 ${\pm 0.31 \pm 0.11 \times 10^2}$
\\
0.736-- 0.858
& 2.88 ${\pm 0.09 \pm 0.11 \times 10^2}$
& 2.64 ${\pm 0.21 \pm 0.10 \times 10^2}$
& 3.10 ${\pm 0.33 \pm 0.12 \times 10^2}$
& 3.25 ${\pm 0.34 \pm 0.12 \times 10^2}$
& 3.59 ${\pm 0.37 \pm 0.14 \times 10^2}$
& 3.48 ${\pm 0.36 \pm 0.13 \times 10^2}$
\\
0.858-- 1.000
& 2.58 ${\pm 0.10 \pm 0.10 \times 10^2}$
& 2.96 ${\pm 0.27 \pm 0.11 \times 10^2}$
& 2.74 ${\pm 0.36 \pm 0.10 \times 10^2}$
& 3.31 ${\pm 0.40 \pm 0.13 \times 10^2}$
& 2.69 ${\pm 0.38 \pm 0.10 \times 10^2}$
& 2.24 ${\pm 0.34 \pm 0.09 \times 10^2}$
\\
1.00-- 1.17
& 2.66 ${\pm 0.16 \pm 0.10 \times 10^2}$
& 2.99 ${\pm 0.43 \pm 0.11 \times 10^2}$
& 2.21 ${\pm 0.52 \pm 0.08 \times 10^2}$
& 2.12 ${\pm 0.51 \pm 0.08 \times 10^2}$
& 2.70 ${\pm 0.60 \pm 0.10 \times 10^2}$
& 3.00 ${\pm 0.63 \pm 0.12 \times 10^2}$
\\
1.17-- 1.36
& 2.66 ${\pm 0.15 \pm 0.10 \times 10^2}$
& 2.66 ${\pm 0.37 \pm 0.10 \times 10^2}$
& 2.84 ${\pm 0.55 \pm 0.11 \times 10^2}$
& 2.78 ${\pm 0.54 \pm 0.11 \times 10^2}$
& 1.96 ${\pm 0.48 \pm 0.08 \times 10^2}$
& 2.57 ${\pm 0.54 \pm 0.10 \times 10^2}$
\\
1.36-- 1.58
& 2.01 ${\pm 0.12 \pm 0.08 \times 10^2}$
& 2.35 ${\pm 0.33 \pm 0.09 \times 10^2}$
& 2.54 ${\pm 0.48 \pm 0.10 \times 10^2}$
& 3.14 ${\pm 0.54 \pm 0.13 \times 10^2}$
& 2.79 ${\pm 0.53 \pm 0.11 \times 10^2}$
& 2.99 ${\pm 0.54 \pm 0.12 \times 10^2}$
\\
1.58-- 1.85
& 2.02 ${\pm 0.11 \pm 0.08 \times 10^2}$
& 1.93 ${\pm 0.27 \pm 0.08 \times 10^2}$
& 1.79 ${\pm 0.37 \pm 0.07 \times 10^2}$
& 1.18 ${\pm 0.31 \pm 0.05 \times 10^2}$
& 1.88 ${\pm 0.40 \pm 0.08 \times 10^2}$
& 2.56 ${\pm 0.46 \pm 0.10 \times 10^2}$
\\
1.85-- 2.15
& 1.87 ${\pm 0.10 \pm 0.08 \times 10^2}$
& 2.03 ${\pm 0.26 \pm 0.08 \times 10^2}$
& 2.34 ${\pm 0.40 \pm 0.10 \times 10^2}$
& 1.22 ${\pm 0.29 \pm 0.05 \times 10^2}$
& 1.84 ${\pm 0.37 \pm 0.08 \times 10^2}$
& 1.71 ${\pm 0.35 \pm 0.07 \times 10^2}$
\\
2.15-- 2.51
& 1.83 ${\pm 0.09 \pm 0.08 \times 10^2}$
& 1.15 ${\pm 0.18 \pm 0.05 \times 10^2}$
& 1.62 ${\pm 0.31 \pm 0.07 \times 10^2}$
& 1.88 ${\pm 0.33 \pm 0.08 \times 10^2}$
& 2.03 ${\pm 0.36 \pm 0.09 \times 10^2}$
& 1.72 ${\pm 0.32 \pm 0.07 \times 10^2}$
\\
2.51-- 2.93
& 1.27 ${\pm 0.07 \pm 0.06 \times 10^2}$
& 1.43 ${\pm 0.19 \pm 0.06 \times 10^2}$
& 8.93 ${\pm 2.10 \pm 0.39 \times 10}$
& 1.56 ${\pm 0.28 \pm 0.07 \times 10^2}$
& 1.52 ${\pm 0.29 \pm 0.07 \times 10^2}$
& 1.21 ${\pm 0.25 \pm 0.05 \times 10^2}$
\\
2.93-- 3.41
& 1.16 ${\pm 0.06 \pm 0.05 \times 10^2}$
& 1.21 ${\pm 0.16 \pm 0.05 \times 10^2}$
& 7.66 ${\pm 1.81 \pm 0.34 \times 10}$
& 1.04 ${\pm 0.21 \pm 0.05 \times 10^2}$
& 1.12 ${\pm 0.23 \pm 0.05 \times 10^2}$
& 9.50 ${\pm 2.07 \pm 0.42 \times 10}$
\\
3.41-- 3.98
& 9.11 ${\pm 0.50 \pm 0.41 \times 10}$
& 9.49 ${\pm 1.32 \pm 0.43 \times 10}$
& 9.55 ${\pm 1.87 \pm 0.43 \times 10}$
& 8.95 ${\pm 1.83 \pm 0.41 \times 10}$
& 7.26 ${\pm 1.71 \pm 0.33 \times 10}$
& 1.09 ${\pm 0.21 \pm 0.05 \times 10^2}$
\\
3.98-- 4.64
& 7.43 ${\pm 0.42 \pm 0.34 \times 10}$
& 7.20 ${\pm 1.06 \pm 0.33 \times 10}$
& 9.14 ${\pm 1.70 \pm 0.42 \times 10}$
& 8.64 ${\pm 1.66 \pm 0.40 \times 10}$
& 8.65 ${\pm 1.73 \pm 0.40 \times 10}$
& 8.03 ${\pm 1.64 \pm 0.37 \times 10}$
\\
4.64-- 5.41
& 6.32 ${\pm 0.36 \pm 0.30 \times 10}$
& 6.05 ${\pm 0.90 \pm 0.29 \times 10}$
& 6.49 ${\pm 1.33 \pm 0.31 \times 10}$
& 6.05 ${\pm 1.29 \pm 0.29 \times 10}$
& 6.83 ${\pm 1.42 \pm 0.32 \times 10}$
& 4.02 ${\pm 1.08 \pm 0.19 \times 10}$
\\
5.41-- 6.31
& 4.57 ${\pm 0.28 \pm 0.22 \times 10}$
& 4.49 ${\pm 0.72 \pm 0.22 \times 10}$
& 5.10 ${\pm 1.09 \pm 0.24 \times 10}$
& 4.47 ${\pm 1.03 \pm 0.21 \times 10}$
& 5.09 ${\pm 1.14 \pm 0.24 \times 10}$
& 4.68 ${\pm 1.07 \pm 0.22 \times 10}$
\\
6.31-- 7.36
& 3.54 ${\pm 0.06 \pm 0.17 \times 10}$
& 3.67 ${\pm 0.17 \pm 0.18 \times 10}$
& 3.59 ${\pm 0.24 \pm 0.18 \times 10}$
& 3.74 ${\pm 0.24 \pm 0.18 \times 10}$
& 3.53 ${\pm 0.24 \pm 0.17 \times 10}$
& 3.43 ${\pm 0.24 \pm 0.17 \times 10}$
\\
7.36-- 8.58
& 2.59 ${\pm 0.05 \pm 0.13 \times 10}$
& 2.69 ${\pm 0.13 \pm 0.13 \times 10}$
& 2.63 ${\pm 0.18 \pm 0.13 \times 10}$
& 3.01 ${\pm 0.19 \pm 0.15 \times 10}$
& 2.38 ${\pm 0.18 \pm 0.12 \times 10}$
& 2.49 ${\pm 0.18 \pm 0.12 \times 10}$
\\
8.58-- 10.0
& 1.95 ${\pm 0.04 \pm 0.10 \times 10}$
& 2.06 ${\pm 0.10 \pm 0.11 \times 10}$
& 1.86 ${\pm 0.14 \pm 0.09 \times 10}$
& 2.05 ${\pm 0.15 \pm 0.10 \times 10}$
& 1.88 ${\pm 0.15 \pm 0.10 \times 10}$
& 2.03 ${\pm 0.15 \pm 0.10 \times 10}$
\\
10.0-- 11.7
& 1.38 ${\pm 0.03 \pm 0.07 \times 10}$
& 1.52 ${\pm 0.08 \pm 0.08 \times 10}$
& 1.30 ${\pm 0.11 \pm 0.07 \times 10}$
& 1.38 ${\pm 0.11 \pm 0.07 \times 10}$
& 1.50 ${\pm 0.12 \pm 0.08 \times 10}$
& 1.42 ${\pm 0.12 \pm 0.07 \times 10}$
\\
11.7-- 13.6
& 1.03 ${\pm 0.02 \pm 0.05 \times 10}$
& 9.22 ${\pm 0.60 \pm 0.48}$
& 9.48 ${\pm 0.87 \pm 0.49}$
& 9.95 ${\pm 0.89 \pm 0.51}$
& 9.88 ${\pm 0.93 \pm 0.51}$
& 8.98 ${\pm 0.87 \pm 0.46}$
\\
13.6-- 15.8
& 7.21 ${\pm 0.19 \pm 0.37}$
& 7.48 ${\pm 0.50 \pm 0.39}$
& 7.23 ${\pm 0.70 \pm 0.37}$
& 6.59 ${\pm 0.67 \pm 0.34}$
& 6.60 ${\pm 0.70 \pm 0.34}$
& 7.10 ${\pm 0.71 \pm 0.37}$
\\
15.8-- 18.5
& 4.72 ${\pm 0.14 \pm 0.24}$
& 4.45 ${\pm 0.36 \pm 0.23}$
& 4.91 ${\pm 0.53 \pm 0.25}$
& 4.52 ${\pm 0.51 \pm 0.23}$
& 4.12 ${\pm 0.51 \pm 0.21}$
& 4.42 ${\pm 0.52 \pm 0.23}$
\\
18.5-- 21.5
& 3.31 ${\pm 0.11 \pm 0.17}$
& 3.33 ${\pm 0.29 \pm 0.17}$
& 3.37 ${\pm 0.41 \pm 0.17}$
& 2.67 ${\pm 0.37 \pm 0.14}$
& 2.50 ${\pm 0.37 \pm 0.13}$
& 2.90 ${\pm 0.39 \pm 0.15}$
\\
      \hline
    \end{tabular}
}
}
\end{table}
\clearpage
\begin{table}
\tiny{
  \caption
{Energy spectrum of protons at the small atmospheric depths measured in BESS-2000 (2).}
  \label{tab:sump2b}
\rotatebox{90}{
    \begin{tabular}{cllllll}
      \hline
      \begin{tabular}{@{}c@{}}       
      Energy range\\
      (GeV nucleon$^{-1}$)
      \end{tabular}
      & 
      \multicolumn{6}{c}{
      \begin{tabular}{@{}c@{}}       
	Flux$\pm~\Delta$Flux$_{\rm sta}
	\pm~\Delta$Flux$_{\rm sys}$\\
	(m$^{-2}$sr$^{-1}$s$^{-1}$GeV$^{-1}$nucleon)
      \end{tabular}
      }
      \\
      \cline{2-7}
      &       \multicolumn{6}{c}{atmospheric depth range [mean] (g/cm$^2$)}\\
      &  \multicolumn{1}{c}{10.0--12.0}
      &  \multicolumn{1}{c}{14.0--17.0} 
      &  \multicolumn{1}{c}{17.0--20.0}   
      &  \multicolumn{1}{c}{20.0--24.0}   
      &  \multicolumn{1}{c}{24.0--30.0}  
      &  \multicolumn{1}{c}{30.0--37.0}   \\
      & \multicolumn{1}{c}{[11.1]} & \multicolumn{1}{c}{[16.1]} 
      & \multicolumn{1}{c}{[18.6]}& \multicolumn{1}{c}{[21.7]}
      & \multicolumn{1}{c}{[27.2]}& \multicolumn{1}{c}{[33.6]}\\
      \hline 
0.185-- 0.215
& 5.54 ${\pm 0.51 \pm 0.21 \times 10^2}$
& 6.29 ${\pm 0.56 \pm 0.24 \times 10^2}$
& 7.16 ${\pm 0.63 \pm 0.27 \times 10^2}$
& 6.95 ${\pm 0.52 \pm 0.26 \times 10^2}$
& 7.89 ${\pm 0.55 \pm 0.30 \times 10^2}$
& 9.11 ${\pm 0.59 \pm 0.34 \times 10^2}$
\\
0.215-- 0.251
& 5.43 ${\pm 0.46 \pm 0.21 \times 10^2}$
& 6.69 ${\pm 0.53 \pm 0.25 \times 10^2}$
& 6.64 ${\pm 0.55 \pm 0.25 \times 10^2}$
& 7.70 ${\pm 0.50 \pm 0.29 \times 10^2}$
& 7.86 ${\pm 0.50 \pm 0.30 \times 10^2}$
& 8.97 ${\pm 0.53 \pm 0.34 \times 10^2}$
\\
0.251-- 0.293
& 4.97 ${\pm 0.41 \pm 0.19 \times 10^2}$
& 5.14 ${\pm 0.43 \pm 0.20 \times 10^2}$
& 7.18 ${\pm 0.53 \pm 0.27 \times 10^2}$
& 6.52 ${\pm 0.42 \pm 0.25 \times 10^2}$
& 7.92 ${\pm 0.46 \pm 0.30 \times 10^2}$
& 7.57 ${\pm 0.45 \pm 0.29 \times 10^2}$
\\
0.293-- 0.341
& 5.10 ${\pm 0.39 \pm 0.19 \times 10^2}$
& 5.18 ${\pm 0.40 \pm 0.20 \times 10^2}$
& 5.80 ${\pm 0.44 \pm 0.22 \times 10^2}$
& 5.99 ${\pm 0.38 \pm 0.23 \times 10^2}$
& 6.46 ${\pm 0.39 \pm 0.25 \times 10^2}$
& 6.67 ${\pm 0.39 \pm 0.25 \times 10^2}$
\\
0.341-- 0.398
& 4.36 ${\pm 0.34 \pm 0.17 \times 10^2}$
& 5.69 ${\pm 0.40 \pm 0.22 \times 10^2}$
& 5.07 ${\pm 0.39 \pm 0.19 \times 10^2}$
& 5.43 ${\pm 0.34 \pm 0.21 \times 10^2}$
& 5.60 ${\pm 0.34 \pm 0.21 \times 10^2}$
& 5.96 ${\pm 0.35 \pm 0.23 \times 10^2}$
\\
0.398-- 0.464
& 3.50 ${\pm 0.29 \pm 0.13 \times 10^2}$
& 4.05 ${\pm 0.32 \pm 0.16 \times 10^2}$
& 4.61 ${\pm 0.36 \pm 0.18 \times 10^2}$
& 4.70 ${\pm 0.30 \pm 0.18 \times 10^2}$
& 5.04 ${\pm 0.31 \pm 0.19 \times 10^2}$
& 5.56 ${\pm 0.33 \pm 0.21 \times 10^2}$
\\
0.464-- 0.541
& 3.99 ${\pm 0.30 \pm 0.15 \times 10^2}$
& 4.07 ${\pm 0.32 \pm 0.16 \times 10^2}$
& 4.24 ${\pm 0.34 \pm 0.16 \times 10^2}$
& 4.12 ${\pm 0.28 \pm 0.16 \times 10^2}$
& 4.61 ${\pm 0.29 \pm 0.18 \times 10^2}$
& 4.91 ${\pm 0.30 \pm 0.19 \times 10^2}$
\\
0.541-- 0.631
& 3.26 ${\pm 0.28 \pm 0.12 \times 10^2}$
& 3.76 ${\pm 0.31 \pm 0.14 \times 10^2}$
& 3.16 ${\pm 0.29 \pm 0.12 \times 10^2}$
& 4.31 ${\pm 0.29 \pm 0.16 \times 10^2}$
& 4.36 ${\pm 0.29 \pm 0.17 \times 10^2}$
& 4.85 ${\pm 0.30 \pm 0.18 \times 10^2}$
\\
0.631-- 0.736
& 3.48 ${\pm 0.30 \pm 0.13 \times 10^2}$
& 3.64 ${\pm 0.31 \pm 0.14 \times 10^2}$
& 3.66 ${\pm 0.33 \pm 0.14 \times 10^2}$
& 3.98 ${\pm 0.29 \pm 0.15 \times 10^2}$
& 3.58 ${\pm 0.27 \pm 0.14 \times 10^2}$
& 3.68 ${\pm 0.27 \pm 0.14 \times 10^2}$
\\
0.736-- 0.858
& 3.27 ${\pm 0.31 \pm 0.12 \times 10^2}$
& 3.76 ${\pm 0.35 \pm 0.14 \times 10^2}$
& 3.49 ${\pm 0.35 \pm 0.13 \times 10^2}$
& 3.46 ${\pm 0.29 \pm 0.13 \times 10^2}$
& 3.66 ${\pm 0.30 \pm 0.14 \times 10^2}$
& 3.85 ${\pm 0.30 \pm 0.15 \times 10^2}$
\\
0.858-- 1.000
& 3.02 ${\pm 0.35 \pm 0.11 \times 10^2}$
& 2.69 ${\pm 0.34 \pm 0.10 \times 10^2}$
& 2.76 ${\pm 0.36 \pm 0.10 \times 10^2}$
& 2.98 ${\pm 0.32 \pm 0.11 \times 10^2}$
& 3.51 ${\pm 0.34 \pm 0.13 \times 10^2}$
& 2.98 ${\pm 0.31 \pm 0.11 \times 10^2}$
\\
1.00-- 1.17
& 1.88 ${\pm 0.44 \pm 0.07 \times 10^2}$
& 2.67 ${\pm 0.54 \pm 0.10 \times 10^2}$
& 2.92 ${\pm 0.60 \pm 0.11 \times 10^2}$
& 2.14 ${\pm 0.43 \pm 0.08 \times 10^2}$
& 3.02 ${\pm 0.50 \pm 0.12 \times 10^2}$
& 2.69 ${\pm 0.47 \pm 0.10 \times 10^2}$
\\
1.17-- 1.36
& 2.51 ${\pm 0.47 \pm 0.10 \times 10^2}$
& 2.38 ${\pm 0.48 \pm 0.09 \times 10^2}$
& 2.92 ${\pm 0.55 \pm 0.11 \times 10^2}$
& 3.08 ${\pm 0.47 \pm 0.12 \times 10^2}$
& 2.73 ${\pm 0.44 \pm 0.11 \times 10^2}$
& 2.44 ${\pm 0.42 \pm 0.10 \times 10^2}$
\\
1.36-- 1.58
& 3.25 ${\pm 0.50 \pm 0.13 \times 10^2}$
& 2.06 ${\pm 0.41 \pm 0.08 \times 10^2}$
& 1.35 ${\pm 0.35 \pm 0.05 \times 10^2}$
& 3.16 ${\pm 0.45 \pm 0.13 \times 10^2}$
& 3.29 ${\pm 0.45 \pm 0.13 \times 10^2}$
& 2.61 ${\pm 0.40 \pm 0.10 \times 10^2}$
\\
1.58-- 1.85
& 1.59 ${\pm 0.32 \pm 0.07 \times 10^2}$
& 1.83 ${\pm 0.36 \pm 0.07 \times 10^2}$
& 2.31 ${\pm 0.42 \pm 0.09 \times 10^2}$
& 1.84 ${\pm 0.32 \pm 0.08 \times 10^2}$
& 1.70 ${\pm 0.30 \pm 0.07 \times 10^2}$
& 2.07 ${\pm 0.33 \pm 0.08 \times 10^2}$
\\
1.85-- 2.15
& 2.22 ${\pm 0.36 \pm 0.09 \times 10^2}$
& 1.64 ${\pm 0.31 \pm 0.07 \times 10^2}$
& 1.86 ${\pm 0.35 \pm 0.08 \times 10^2}$
& 1.54 ${\pm 0.27 \pm 0.06 \times 10^2}$
& 2.56 ${\pm 0.34 \pm 0.11 \times 10^2}$
& 2.29 ${\pm 0.32 \pm 0.10 \times 10^2}$
\\
2.15-- 2.51
& 1.67 ${\pm 0.29 \pm 0.07 \times 10^2}$
& 1.46 ${\pm 0.28 \pm 0.06 \times 10^2}$
& 1.03 ${\pm 0.24 \pm 0.04 \times 10^2}$
& 1.89 ${\pm 0.28 \pm 0.08 \times 10^2}$
& 1.62 ${\pm 0.25 \pm 0.07 \times 10^2}$
& 1.54 ${\pm 0.25 \pm 0.07 \times 10^2}$
\\
2.51-- 2.93
& 1.10 ${\pm 0.22 \pm 0.05 \times 10^2}$
& 1.17 ${\pm 0.23 \pm 0.05 \times 10^2}$
& 1.47 ${\pm 0.27 \pm 0.06 \times 10^2}$
& 1.11 ${\pm 0.20 \pm 0.05 \times 10^2}$
& 1.39 ${\pm 0.22 \pm 0.06 \times 10^2}$
& 1.22 ${\pm 0.20 \pm 0.05 \times 10^2}$
\\
2.93-- 3.41
& 1.49 ${\pm 0.23 \pm 0.07 \times 10^2}$
& 1.08 ${\pm 0.20 \pm 0.05 \times 10^2}$
& 6.33 ${\pm 1.63 \pm 0.28 \times 10}$
& 1.04 ${\pm 0.18 \pm 0.05 \times 10^2}$
& 9.89 ${\pm 1.70 \pm 0.44 \times 10}$
& 8.43 ${\pm 1.57 \pm 0.37 \times 10}$
\\
3.41-- 3.98
& 9.39 ${\pm 1.71 \pm 0.43 \times 10}$
& 1.06 ${\pm 0.19 \pm 0.05 \times 10^2}$
& 1.02 ${\pm 0.19 \pm 0.05 \times 10^2}$
& 1.69 ${\pm 0.21 \pm 0.08 \times 10^2}$
& 1.03 ${\pm 0.16 \pm 0.05 \times 10^2}$
& 1.00 ${\pm 0.16 \pm 0.05 \times 10^2}$
\\
3.98-- 4.64
& 6.44 ${\pm 1.31 \pm 0.30 \times 10}$
& 6.84 ${\pm 1.40 \pm 0.32 \times 10}$
& 9.37 ${\pm 1.71 \pm 0.43 \times 10}$
& 6.59 ${\pm 1.20 \pm 0.30 \times 10}$
& 6.46 ${\pm 1.18 \pm 0.30 \times 10}$
& 5.17 ${\pm 1.05 \pm 0.24 \times 10}$
\\
4.64-- 5.41
& 6.68 ${\pm 1.24 \pm 0.32 \times 10}$
& 4.90 ${\pm 1.09 \pm 0.23 \times 10}$
& 7.24 ${\pm 1.39 \pm 0.34 \times 10}$
& 5.28 ${\pm 1.00 \pm 0.25 \times 10}$
& 6.66 ${\pm 1.11 \pm 0.31 \times 10}$
& 7.21 ${\pm 1.15 \pm 0.34 \times 10}$
\\
5.41-- 6.31
& 4.14 ${\pm 0.90 \pm 0.20 \times 10}$
& 6.71 ${\pm 1.19 \pm 0.32 \times 10}$
& 3.21 ${\pm 0.86 \pm 0.15 \times 10}$
& 3.87 ${\pm 0.79 \pm 0.19 \times 10}$
& 3.48 ${\pm 0.74 \pm 0.17 \times 10}$
& 3.96 ${\pm 0.79 \pm 0.19 \times 10}$
\\
6.31-- 7.36
& 3.52 ${\pm 0.21 \pm 0.17 \times 10}$
& 3.69 ${\pm 0.23 \pm 0.18 \times 10}$
& 2.96 ${\pm 0.21 \pm 0.15 \times 10}$
& 3.60 ${\pm 0.20 \pm 0.18 \times 10}$
& 2.91 ${\pm 0.18 \pm 0.14 \times 10}$
& 3.19 ${\pm 0.18 \pm 0.16 \times 10}$
\\
7.36-- 8.58
& 2.62 ${\pm 0.17 \pm 0.13 \times 10}$
& 2.45 ${\pm 0.17 \pm 0.12 \times 10}$
& 2.37 ${\pm 0.17 \pm 0.12 \times 10}$
& 2.44 ${\pm 0.15 \pm 0.12 \times 10}$
& 2.59 ${\pm 0.15 \pm 0.13 \times 10}$
& 2.26 ${\pm 0.14 \pm 0.11 \times 10}$
\\
8.58-- 10.0
& 1.88 ${\pm 0.13 \pm 0.10 \times 10}$
& 1.82 ${\pm 0.13 \pm 0.09 \times 10}$
& 1.98 ${\pm 0.14 \pm 0.10 \times 10}$
& 1.88 ${\pm 0.12 \pm 0.10 \times 10}$
& 1.90 ${\pm 0.12 \pm 0.10 \times 10}$
& 1.60 ${\pm 0.11 \pm 0.08 \times 10}$
\\
10.0-- 11.7
& 1.59 ${\pm 0.11 \pm 0.08 \times 10}$
& 1.18 ${\pm 0.10 \pm 0.06 \times 10}$
& 1.32 ${\pm 0.11 \pm 0.07 \times 10}$
& 1.34 ${\pm 0.09 \pm 0.07 \times 10}$
& 1.22 ${\pm 0.09 \pm 0.06 \times 10}$
& 1.12 ${\pm 0.08 \pm 0.06 \times 10}$
\\
11.7-- 13.6
& 9.48 ${\pm 0.80 \pm 0.49}$
& 1.07 ${\pm 0.09 \pm 0.06 \times 10}$
& 9.70 ${\pm 0.87 \pm 0.50}$
& 8.69 ${\pm 0.69 \pm 0.45}$
& 8.58 ${\pm 0.68 \pm 0.44}$
& 8.95 ${\pm 0.70 \pm 0.46}$
\\
13.6-- 15.8
& 7.08 ${\pm 0.64 \pm 0.37}$
& 6.11 ${\pm 0.61 \pm 0.32}$
& 5.15 ${\pm 0.59 \pm 0.27}$
& 6.40 ${\pm 0.55 \pm 0.33}$
& 5.82 ${\pm 0.52 \pm 0.30}$
& 5.17 ${\pm 0.49 \pm 0.27}$
\\
15.8-- 18.5
& 4.68 ${\pm 0.48 \pm 0.24}$
& 4.81 ${\pm 0.50 \pm 0.25}$
& 5.27 ${\pm 0.55 \pm 0.27}$
& 4.23 ${\pm 0.41 \pm 0.22}$
& 4.42 ${\pm 0.42 \pm 0.23}$
& 4.97 ${\pm 0.44 \pm 0.26}$
\\
18.5-- 21.5
& 2.83 ${\pm 0.35 \pm 0.15}$
& 3.23 ${\pm 0.38 \pm 0.17}$
& 3.54 ${\pm 0.42 \pm 0.18}$
& 3.42 ${\pm 0.34 \pm 0.18}$
& 2.44 ${\pm 0.29 \pm 0.13}$
& 2.57 ${\pm 0.30 \pm 0.13}$
\\
      \hline
    \end{tabular}
}
}
\end{table}

\clearpage

%
%
\clearpage
\clearpage
\clearpage
\clearpage
\clearpage
\clearpage
\clearpage

\begin{thebibliography}{99}

\bibitem{papini} P. Papini, {\it et al.},
Nuovo Cimento 19C (1996) 367.

\bibitem{ams01} J. Alcaraz, {\it et al.},
Phys. Lett. B 490 (2000) 27.

\bibitem{cap98} M. Boezio, {\it et al.},
Astropart. Phys. 19 (2003) 583 and references therein.

\bibitem{kn:prop87} 
S.~Orito, in: Proceedings of ASTROMAG Workshop, 
KEK Report KEK87-19 (1987) 111.

\bibitem{YA88} A.~Yamamoto, {\it et al.}, 
IEEE Trans. Magn. 24 (1988) 1421.

\bibitem{YA94} A.~Yamamoto, {\it et al.},
Adv. Space Res. 14 (1994) 75.

\bibitem{Ajima} Y. Ajima, {\it et al.},
Nucl. Instrum. Methods A 443 (2000) 71.

\bibitem{tofp} Y. Shikaze, {\it et al.},
Nucl. Instrum. Methods A 455 (2000) 596.

\bibitem{asaoka98} Y. Asaoka, {\it et al.},
Nucl. Instrum. Methods A 416 (1998) 236.

\bibitem{sanuki-phe} T. Sanuki, {\it et al.},
Astrophys. J. 545 (2000) 1135.

\bibitem{abe-phe} K. Abe, {\it et al.},
Phys. Lett. B 564 (2003) 8.

\bibitem{haino-phe} S. Haino, {\it et al.}, 
Phys. Lett. B 594 (2004) 35.

\bibitem{motoki-mu} M. Motoki, {\it et al.},
Astropart. Phys. 19 (2003) 113.

\bibitem{norikura-ppb} T. Sanuki, {\it et al.},
Phys. Lett. B 577 (2003) 10.

\bibitem{YO95} K. Yoshimura, {\it et al.}, 
Phys. Rev. Lett. 75 (1995) 3792; 
A. Moiseev, {\it et al.}, Astrophys. J. 474 (1997) 479.

\bibitem{MA98} H. Matsunaga, {\it et al.}, 
Phys. Rev. Lett. 81 (1998) 4052.

\bibitem{OR00} S. Orito, {\it et al.}, 
Phys. Rev. Lett. 84 (2000) 1078.

\bibitem{MA01} T. Maeno, {\it et al.}, 
Astropart. Phys. 16 (2001) 121.

\bibitem{AS01} Y. Asaoka, {\it et al.}, 
Phys. Rev. Lett. 88 (2002) 051101.

\bibitem{book02} M. Sasaki, {\it et al.},
Nucl. Phys. B (Proc. Suppl.) 113 (2002) 202.

\bibitem{Karimaki} V.~Karimaki, 
Comput. Phys. Commun. 69 (1992) 133.

\bibitem{haino-nim} S. Haino, {\it et al.},
Nucl. Instrum. Methods A 518 (2004) 167.

\bibitem{beamp} Y. Asaoka, {\it et al.},
Nucl. Instrum. Methods A 489 (2001) 170.

\bibitem{NEUT}
http://odysseus.uchicago.edu/NeutronMonitor/:
University of Chicago, 
``National Science Foundation Grant ATM-9613963''.

\bibitem{geant} R. Brun, {\it et al.},  
GEANT--Detector Description and Simulation Tool, 
CERN Program Library, Long Write up W5013. 

\bibitem{atmnc} M. Honda, {\it et al.}, 
Phys. Rev. D70 (2004) 043008. 

\bibitem{seo} E.S. Seo, {\it et al.},
Proceedings of 25th International Cosmic-Ray Conference, Durban, 3 (1997) 337.

\bibitem{engelm} J.J. Engelmann, {\it et al.},
Astron. Astrophys. 233 (1990) 96.

\bibitem{sullivan1971} J.D. Sullivan, 
Nucl. Instrum. Methods 95 (1971) 5.

\bibitem{wang} J. Z. Wang, {\it et al.},
Astrophys. J. 564 (2002) 244.

\bibitem{FFA} L.J. Gleeson, W.I. Axford,
Astrophys. J. 154 (1968) 1011.

\bibitem{Myers} Z.D. Myers, E.S. Seo,
Adv. Space Res. 35 (2005) 151.

\bibitem{bieber} J.W. Bieber {\it et al.}, Phys. Rev. Lett. 83 (1999) 674.

\bibitem{miyake} S. Miyake and S. Yanagita, 
Proceedings of 29th International Cosmic-Ray Conference, Pune, 2 (2005) 203; 
astro-ph/0610777; S. Miyake and S. Yanagita, private communication.

\end{thebibliography}
\end{document}